\documentclass[journal]{IEEEtran}
\usepackage{cite}

\ifCLASSINFOpdf
  \usepackage[pdftex]{graphicx}
\else
\fi
\usepackage[caption=false, font=footnotesize]{subfig}

\usepackage{amsmath}
\usepackage{amsmath,mleftright}
\usepackage{amssymb}
\usepackage{amsthm}
\usepackage{xparse}
\NewDocumentCommand{\evalat}{sO{\big}mm}{%
  \IfBooleanTF{#1}
   {\mleft. #3 \mright|_{#4}}
   {#3#2|_{#4}}%
}
\DeclareMathOperator*{\argmin}{argmin}

\usepackage{url}
\usepackage{array,multirow}

\begin{document}
%
\title{Sensor-Aided Learning for Wi-Fi Positioning\\ with Beacon Channel State Information}
%
%
%

\author{Jeongsik~Choi
\thanks{J.~Choi is with the Intel Labs, Intel Corporation, Santa Clara, CA 95054, USA (e-mail: jeongsik.choi@intel.com).}
}
\maketitle

\begin{abstract}
Because each indoor site has its own radio propagation characteristics, a site survey process is essential to optimize a Wi-Fi ranging strategy for range-based positioning solutions.
This paper studies an unsupervised learning technique that autonomously investigates the characteristics of the surrounding environment using sensor data accumulated while users use a positioning application.
Using the collected sensor data, the device trajectory can be regenerated, and a Wi-Fi ranging module is trained to make the shape of the estimated trajectory using Wi-Fi similar to that obtained from sensors.
In this process, the ranging module learns the way to identify the channel conditions from each Wi-Fi access point (AP) and produce ranging results accordingly.
Furthermore, we collect the channel state information (CSI) from beacon frames and evaluate the benefit of using CSI in addition to received signal strength (RSS) measurements.
When CSI is available, the ranging module can identify more diverse channel conditions from each AP, and thus more precise positioning results can be achieved.
The effectiveness of the proposed learning technique is verified using a real-time positioning application implemented on a PC platform. 
\end{abstract}


%
\IEEEpeerreviewmaketitle

\section{Introduction}

\IEEEPARstart{W}{ith} the advent of various types of mobile devices, interest in location-based services has greatly increased in recent decades.
Precise and reliable positioning is a key technology that enhances the end-user experience and creates new business opportunities.
To achieve high-quality positioning results, many efforts have been introduced in the literature~\cite{4796924, correa_sen_17, 8409950}.
In particular, there is a great demand for positioning solutions that rely only on built-in components of mobile devices and do not require the installation of additional infrastructure.

In this context, Wi-Fi has been widely used for locating mobile devices indoor.
Because Wi-Fi access points (APs) are easy and cost-effective to deploy, many indoor sites already have dense APs that can be used for locating mobile devices.
Without being associated with an AP, mobile devices can listen to beacon frames broadcast from nearby APs, thereby having received signal strength (RSS).
For this reason, RSS has been used as a primary source for positioning solutions that are based on either trilateration methods~\cite{Wang2003AnIW,5425237,7935650,8839041} or fingerprinting methods~\cite{832252,1047316,horus05,Liu2012}.
However, RSS is affected by many factors, such as small-scale fading and body shadowing, resulting in the degradation of positioning quality.

For this reason, many studies have focused on channel state information (CSI), which is available with commodity Wi-Fi devices, such as Intel IWL5300~\cite{csi_tool_intel} and the Atheros series~\cite{csi_tool_atheros}.
Because CSI provides fine-grained information about the propagation channel, it can improve the positioning performance in various ways, such as identifying propagation channel conditions~\cite{6646249, 6848217, 7130677, 7218588, 8166758, s18114057, Liu2019, 8320781}, replacing RSS from fingerprinting methods~\cite{6289200, 7438932, 8027020, 8187642, 8485917}, and estimating the angle of arrival (AoA) or time of flight (ToF) of the wireless signal~\cite{10.5555/2482626.2482635, 10.1145/2462456.2464463, 10.1145/2639108.2639142, 10.1145/2785956.2787487, Ahmed2018, 8718525}.
Nevertheless, there have been some implementation limitations because these CSI tools can capture the CSI of only high throughput (HT) packets. 
Transmitters must transmit HT packets, and simultaneously collecting CSI from multiple APs is challenging.

To enhance CSI measurement capability, Intel has developed a new CSI tool for the latest Wi-Fi chipsets, such as Wireless-AC9260/9560 and Wi-Fi 6 AX200/201 series.
This tool is designed to capture the CSI of any orthogonal frequency division multiplexing (OFDM) format defined in the IEEE 802.11a/g/n/ac/ax standards, including legacy OFDM, HT, and very high throughput (VHT) formats.
Consequently, this tool can capture the CSI of a beacon frame that is broadcast with a legacy OFDM format for backward compatibility.
Without modifying any setting of existing APs, the CSI from the nearby APs can be simultaneously collected using this tool in the same way as the device collects RSS from beacon frames.

Although the CSI of the beacon frame is available, there are still challenges. (i) The transmission bandwidth of the beacon frame is 20~MHz, which may not sufficient to obtain high resolution multipath propagation profiles from transmitters. This can degrade the performance of existing CSI-based processes, such as identifying channel conditions or estimating the AoA.
(ii) APs and devices are not synchronized for beacon transmission and reception; thus, distance estimation based on the ToF of the beacon frame can produce incorrect results.
(iii) The amount of available CSI is limited because beacons frames are broadcast with a predefined interval (e.g., 100~ms).
Thus, it is difficult to monitor subtle changes in the channel over a short time.

Despite these challenges, the CSI of the beacon frame is still useful for improving the positioning performance because it can compensate for the unstable fluctuation nature of RSS using wideband information about the channel.
In addition, the presence of a dominant path among the multipath components can be inferred from the frequency selectivity of the CSI.
One issue is that the indoor propagation channel is too diverse and complex to be explained using a single model.
For this reason, the method of extracting useful features from the CSI of the beacon frame and the method to effectively exploit the extracted features to achieve accurate positioning results can widely vary from indoor site to site.

In this situation, machine learning techniques can play an important role in minimizing the time and effort required for analyzing the characteristics of each indoor site.
Many studies have demonstrated that various machine learning architectures can be successfully deployed in the positioning field to minimize human intervention and improve performance~\cite{7536949, 7438932, 10.1145/3194554.3194594, 8027020, 8533766, intel19}.
Nevertheless, most previous studies have relied on supervised learning techniques that require the collection of ground truth data, such as the true x- and y-coordinates of the device. 
Therefore, the training data have been manually obtained by collecting data at preset marker locations in indoor sites, or using high-precision equipment, such as LIDAR (Light Detection and Ranging)~\cite{intel19}.

In this study, we consider a positioning solution based on Wi-Fi ranging and built-in sensors of mobile devices.
This work was motivated by a sensor fusion technique called pedestrian dead reckoning (PDR), which has been widely deployed in positioning solutions~\cite{Liu2012,Tian2014,Diaz2015,6987239,6971168,7935650,Wang2018,8385119,8756098,8756276,Xu2019, 9125898}. 
The PDR technique utilizes the accelerometer and gyroscope readings to estimate the trajectory of a device, by counting the number of steps of a user holding the device and estimating the heading direction.
Because the PDR technique can provide an accurate trajectory of the device, we can appropriately process the PDR output and exploit it in the training phase. 

\begin{figure}
    \centering
    \includegraphics[width=0.45\textwidth]{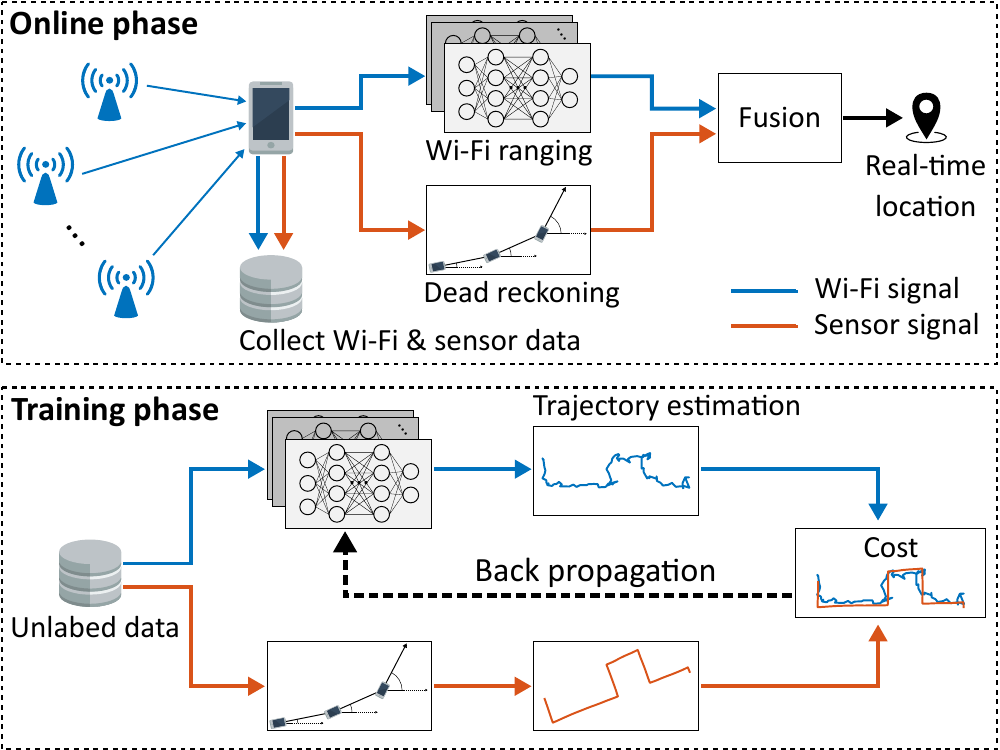}
    \caption{Overview of the proposed sensor-aided learning framework.}
    \label{fig_paper_overview}
\end{figure}

Fig.~\ref{fig_paper_overview} shows an overview of the proposed learning framework.
During the online phase, a positioning application fuses the Wi-Fi ranging and PDR outputs to obtain the real-time location of the device.
At the same time, the application collects Wi-Fi and sensor data.
In the training phase, the collected data are used to separately estimate the trajectory of the device, and a cost function is designed to measure the similarity between the two estimated trajectories.
Because the cost function does not include any ground truth information, training data can be collected in a crowdsourcing manner, and the accuracy of the ranging module can be improved as the data accumulate.

The contributions of this study are summarized as follows:
\begin{enumerate}
    \item [1)] We deployed neural networks (NNs) to estimate the distance from nearby APs using any available information and designed a cost function that exploits Wi-Fi and sensor data generated inside devices.
    Therefore, the proposed method significantly reduces human intervention for collecting training data. 
    \item [2)] In addition to RSS, the CSI of the beacon frame was collected using a commodity Wi-Fi chipset, and the benefit of using the CSI was verified. To best of our knowledge, this is the first work that uses the new CSI tool with the latest Intel Wi-Fi chipset.
    \item [3)] To demonstrate the effectiveness of the proposed method, we implemented a real-time positioning application for a PC platform and conducted extensive experiments in a  large-scale indoor office environment with 59 Wi-Fi APs. A real-time demo video is available online~\cite{demo_video}.
\end{enumerate}

\emph{Organization}: The remainder of this paper is organized as follows:
In Section~II, the background of CSI and sensor fusion technique are discussed. In Section~III, the ranging module using NNs and positioning techniques to estimate the location of the device using ranging results is introduced.
In Section~IV, a new cost function designed to train the ranging module with sensor information is given. The experimental results are presented in Section~V, and the conclusion is in Section~VI.

\emph{Notation}: $\mathbf A\in\mathbb{R}^{N\times M}$ represents an $N\times M$ real matrix (or vector) where $[\mathbf A]_{(n, m)}$ indicates the $(n, m)$-th element of the matrix.
$\mathbf {I}_{N} \in \mathbb{R}^{N\times N}$ is the identity matrix and $\mathbf A = \mbox{diag}(a_1, ..., a_N) \in \mathbb{R}^{N\times N}$ denotes the diagonal matrix with diagonal elements $a_1, ..., a_N$. 
The transpose and inverse operations are denoted by $(\cdot)^T$ and $(\cdot)^{-1}$, respectively.
The L2-norm of a vector $\mathbf a \in \mathbb{R}^{N \times 1}$ is denoted by $\lVert \mathbf a \rVert = \sqrt{\mathbf{a}^T\mathbf{a}}$, and the expectation operator is denoted by $E[\cdot]$.

\section{Background}

\subsection{Channel State Information}

Under multipath propagation environments, each OFDM sub-carrier experiences a unique distortion.
Thus, the Wi-Fi system performs the channel sounding procedure to measure the channel coefficient of each sub-carrier, called the CSI.

CSI is related to the channel impulse response (CIR), which is expressed by
\begin{equation} \label{3B1}
    h(t) = \sum_{l=0}^{L-1} c_l \delta(t - \tau_l),
\end{equation}
where $\delta(\cdot)$ represents the Dirac delta function, and $L$ denotes the number of multipath components between the transmitter and receiver. The $l$-th multipath component is characterized by a complex channel coefficient and a time delay, which are denoted by $c_l$ and $\tau_l$, respectively.
The frequency response of the sub-carrier $n$ is expressed by
\begin{equation} \label{3B2}
    H(f_n) = \int_{-\infty}^{\infty} h(t)e^{-j2\pi f_n t} dt = \sum_{l=0}^{L-1}c_l e^{-j2\pi f_n \tau_l},
\end{equation}
where $f_n = n\Delta f = \frac{n}{N_{DFT}T_s}$ indicates the baseband frequency of the  sub-carrier~$n$, $N_{DFT}$ is the size of the discrete Fourier transform, and $T_s$ is the sampling rate.
With additive measurement noise, the CSI of the sub-carrier $n$ is expressed by
\begin{equation} \label{3B3}
    \hat{H}(f_n) = H(f_n) + \nu_n,
\end{equation}
where $\nu_n$ is a zero-mean complex random variable.

Because the main interest in this study is the CSI of the beacon frame, parameters for the legacy OFDM format are used, which are given as $N_{DFT}=64$, $\frac{1}{T_s} = 20$~MHz, and $\Delta f = 312.5$~kHz.
Using the new CSI tool mentioned in Introduction, we can collect the complex frequency response of 52 sub-carriers, which are pilot and data sub-carriers with index $n\in \{-26, ..., -1, 1, ..., 26\}$.
When multiple antennas are used for beacon reception, multiple CSI sets are obtained accordingly.
The measured frequency responses are reported with 10-bit resolution for the real and imaginary parts.

\subsection{Pedestrian Dead Reckoning}

This work primarily focuses on a scenario where users use a positioning application on their handheld devices. Therefore, the PDR technique is used to estimate the trajectory of the device.
For other scenarios, such as a robot is used for site survey purposes, other appropriate techniques can be applied instead of the PDR technique.
We briefly summarize the PDR procedure, where the details can be found in~\cite{9125898}.

\begin{figure}
    \centering
    \subfloat[]{\includegraphics[width=0.15\textwidth]{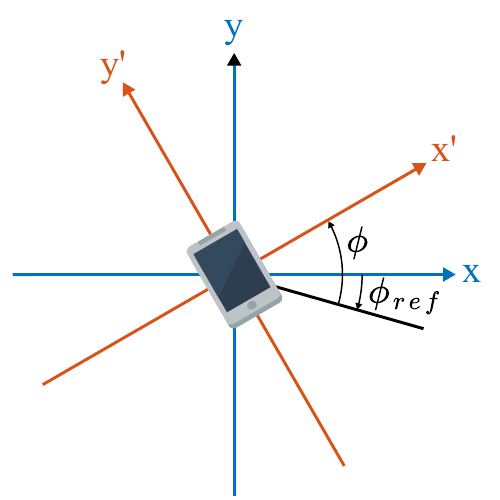}}\hfil
    \subfloat[]{\includegraphics[width=0.30\textwidth]{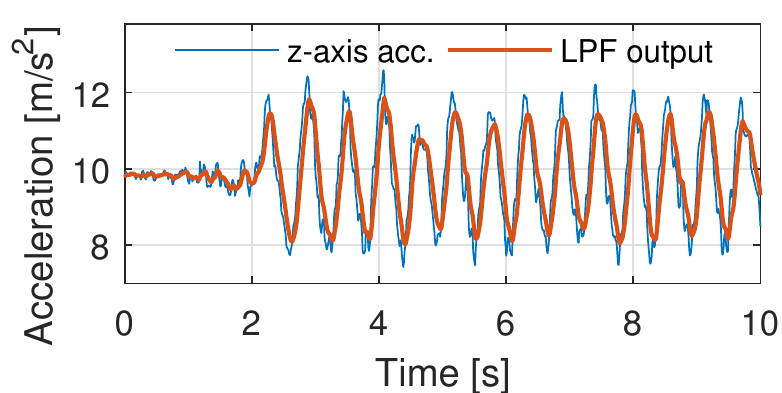}}
    \caption{Two main processes of the PDR module: (a)~Heading estimation and (b)~step detection with the z-axis acceleration.}
    \label{fig_pdr}
\end{figure}

As a first step, the orientation of the device should be obtained.
To this end, many sensor fusion techniques have been proposed in the literature to estimate the orientation using a built-in accelerometer that measures the gravity of the Earth and gyroscope that detects the rotation of the device~\cite{Feng2017, Bernal-Polo2019}.
With the estimated orientation, the two dimensional rotation of the device on an x-y plane of the global coordinate system (GCS) can be obtained, as shown in Fig.~\ref{fig_pdr}(a).
The x-, y-, and z-axes of the GCS point in the East, North and up (ENU) directions relative to the position of the device on the Earth's surfaces whereas x$'$- and y$'$-axes point in right and upward directions of the device, respectively. 
The device is assumed to move along the y$'$-axis, and the two dimensional rotation of the device on the x-y plane is expressed as $\phi$ and is called the heading angle.
Note that a magnetometer is not used in this study because of the distortion of the magnetic field in indoor environment. Therefore $\phi$ represents the angle relative to an arbitrary reference direction denoted as $\phi_{ref}$.

The estimated orientation of the device is also used to transform the local accelerometer readings relative to the GCS.
We denote $\mathbf a = [a_x, a_y, a_z]^T$ as the transformed three-axis acceleration of the device on the GCS.
Among the transformed values, the z-axis component captures an up and down movement pattern of the device, which is generated when the user is walking.
Fig.~\ref{fig_pdr}(b) shows a periodic pattern of the z-axis acceleration and its low pass filter (LPF) output.
A peak followed by a valley can be considered as a step of the user holding the device. 
The step length associated with the detected peak and valley can be estimated using a non-linear step length model as follows~\cite{5646888}:
\begin{equation} \label{3b3}
    \mathcal L = \alpha (a_{z, max} - a_{z, min})^{\frac{1}{4}},
\end{equation}
where $a_{z, max}$ and $a_{z, min}$ represent the peak and valley accelerations, respectively, and $\alpha$ is a constant coefficient.

According to the step detection results, the estimated position of the device at time $t$ is expressed as
\begin{equation} \label{3b4}
    \mathbf{p}(t) = \mathbf p(t-1) + \mathcal L(t) \mathbf u\left(\phi(t)\right),
\end{equation}
where $\mathcal L(t)$ indicates the step length computed using equation~(\ref{3b3}) when a step is detected at time $t$ and 0 if otherwise.
In addition, $\phi(t)$ represents the heading angle estimated at time~$t$, and $\mathbf u(\phi) = [-\sin\phi, \cos\phi]^T$ is the moving direction of the device on the x-y plane of the GCS.
Note that a magnetometer is not used in this study because of the distortion of the magnetic field in indoor environment.

\section{Ranging and Positioning Using Beacon CSI}

\subsection{Assumptions}

We consider a positioning scenario where multiple Wi-Fi APs are installed on an x-y plane of the GCS, for instance, on the same floor of an indoor site.
At each time step, the device scans all Wi-Fi channels used by APs in the vicinity.
By receiving beacon frames broadcast from nearby APs, the device can obtain both RSS and CSI.
Moreover, we assume that the device simultaneously activates two receive antennas for capturing beacon frames. 
Therefore, two sets of RSS and CSI can be obtained by receiving a single beacon frame.

Even within a single channel scanning procedure, the device can capture multiple beacon frames from each AP by increasing the proving time for each Wi-Fi channel.
To address this, we denote $B$ as the number of received beacon frames used for the ranging procedure ($B\geq 1$).
When the device receives fewer than $B$ beacon frames from a certain AP, the ranging results from that AP are not available.
Among the many APs with available ranging results, the positioning module selects up to $N$ APs to estimate the position of the device.

We denote $\mathbf z = [x, y]^T$ as the position of the device on the x-y plane, and $\mathbf z_n^{(k)} = [x_n^{(k)}, y_n^{(k)}]^T$ is the position of the $n$-th selected AP at time step $k$.
The superscript $(\cdot)^{(k)}$ is attached to the position of APs to ensure that the position of $N$ selected APs can change over time step as the device moves.
We assume that the coordinates of all APs are known through a one-time manual effort or an automated method introduced in~\cite{8911824}.

\subsection{CNN-Based Ranging with Beacon CSI}

From the information obtained by receiving beacon frames broadcast from a nearby AP, we can estimate both the distance from the AP and the expected standard deviation of the distance estimate.
The input and output relationship of the ranging module is expressed as a parametric function as follows:
\begin{equation} \label{4b1}
    \mathcal R(\mathcal{X}; \Theta) = \begin{bmatrix} \hat{d} \\ \hat{s}\end{bmatrix},
\end{equation}
where $\mathcal X$ represents the input layer that can include any information obtained from the received beacon frames.
The two outputs denoted by $\hat{d}$ and $\hat{s}$ represent the distance estimate and its standard deviation, respectively.

\begin{figure}
    \centering
    \subfloat[]{\includegraphics[width=0.24\textwidth]{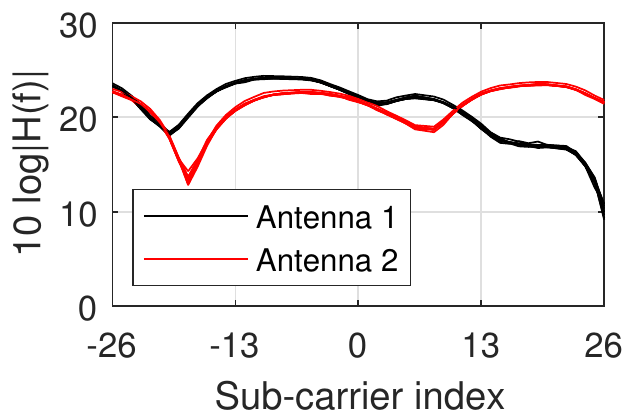}}\hfil\hfil
    \subfloat[]{\includegraphics[width=0.24\textwidth]{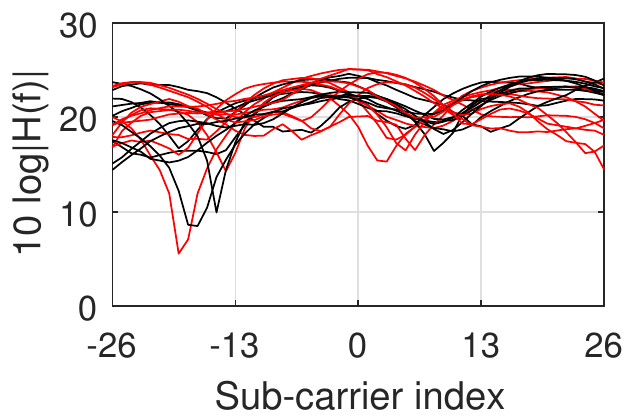}}\hfil\hfil
    \subfloat[]{\includegraphics[width=0.24\textwidth]{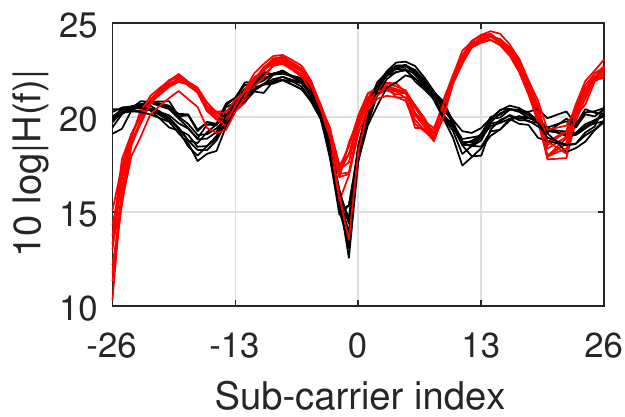}}\hfil\hfil
    \subfloat[]{\includegraphics[width=0.24\textwidth]{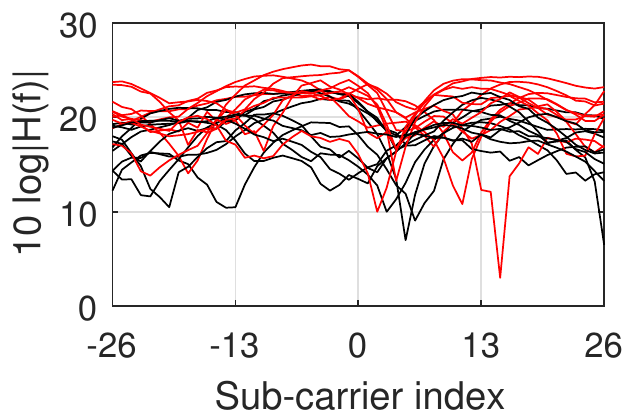}}
    \caption{Amplitude of CSI obtained from 20 beacon frames from an AP for the following scenarios: (a) The device is stationary under a LOS condition, (b) the device is moving under a LOS condition, (c) the device is stationary under an NLOS condition, and (d) the device is moving under an NLOS condition.}
    \label{fig_csi}
\end{figure}

Fig.~\ref{fig_csi} illustrates the amplitude of CSI obtained from 20 consecutive beacon frames transmitted from an AP.
Fig.~\ref{fig_csi}(a) and (c) represent the results when the device is stationary under LOS and NLOS conditions, respectively.
The coherence bandwidth of the channel under the LOS condition is wider than that under the NLOS condition.
When the device is moving, the CSI fluctuates widely over time, as shown in Fig.~\ref{fig_csi}(b) and (d).
Nevertheless, the coherence bandwidth of the CSI of each beacon frame is relatively wide under the LOS condition.

To extract useful features from the CSI and RSS efficiently, we can include all information in the input layer as 
\begin{equation}
    \mathcal X = [\mathcal X_{CSI}, \mathcal X_{RSS}],
\end{equation}
where $\mathcal X_{CSI}$ represents the input related to the CSI. 
For simplicity, we only consider the amplitude of CSI and represent $\mathcal X_{CSI}$ as a two-channel image, where each channel is constructed from $B$ received CSI using an antenna.
Therefore, each channel image is expressed as a two dimensional matrix in $\mathbb{R}^{B\times 52}$, where each row consists of the amplitude of $\hat{H}(f_n)$ for $n\in\{-26, ..., -1, 1, ..., 26\}$.
Furthermore, $\mathcal X_{RSS}$ represent two RSS vectors, each with $B$ RSS measurements obtained using an antenna.

\begin{figure}[]
    \centering
    \subfloat[]{\includegraphics[width=0.24\textwidth]{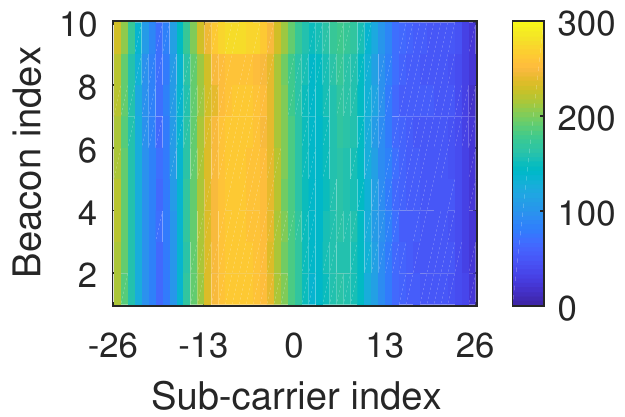}}\hfil\hfil
    \subfloat[]{\includegraphics[width=0.24\textwidth]{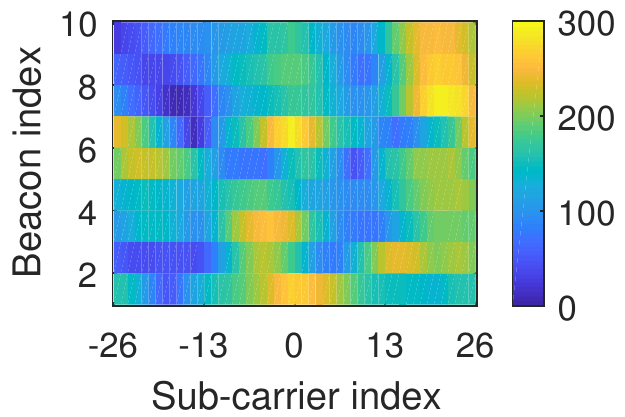}}\hfil\hfil
    \subfloat[]{\includegraphics[width=0.24\textwidth]{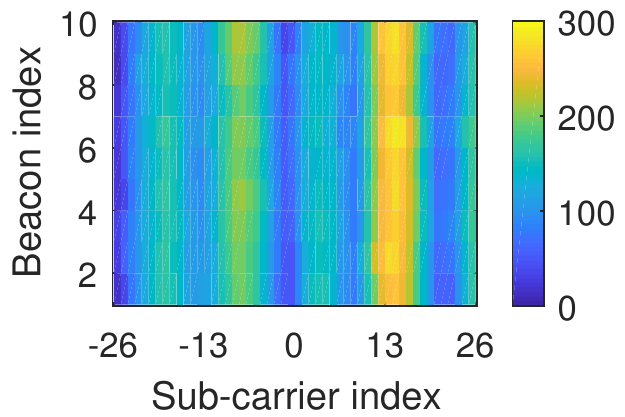}}\hfil\hfil
    \subfloat[]{\includegraphics[width=0.24\textwidth]{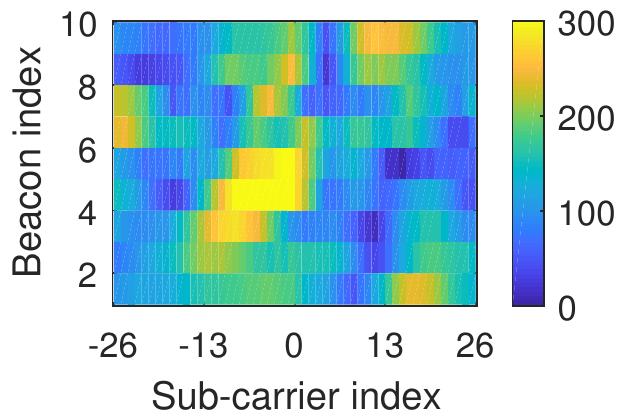}}
    \caption{Examples of one channel CSI input image constructed under: (a) A LOS condition when the device is stationary, (b) a LOS condition when the device is moving, (c) an NLOS condition when the device is stationary, and (d) an NLOS condition when the device is moving.}
    \label{fig_csi_input}
\end{figure}

Fig.~\ref{fig_csi_input} visualizes one channel of CSI input image.
Fig.~\ref{fig_csi_input}(a) and (c) depict the results when the user is stationary under LOS and NLOS conditions, respectively.
Because the measured CSI is stationary over time, a few vertical stripes can be seen in both figures and each strip in Fig.~\ref{fig_csi_input}(a) is wider than each strip in Fig.~\ref{fig_csi_input}(c).
By contrast, Fig.~\ref{fig_csi_input}(b) and (d) show the CSI input image for a scenario when the user is moving under LOS and NLOS conditions, respectively.
Although the CSI fluctuates widely, each row of the input image shows different frequency selective patterns for the two scenarios.

\begin{figure}
    \centering
    \includegraphics[width=0.45\textwidth]{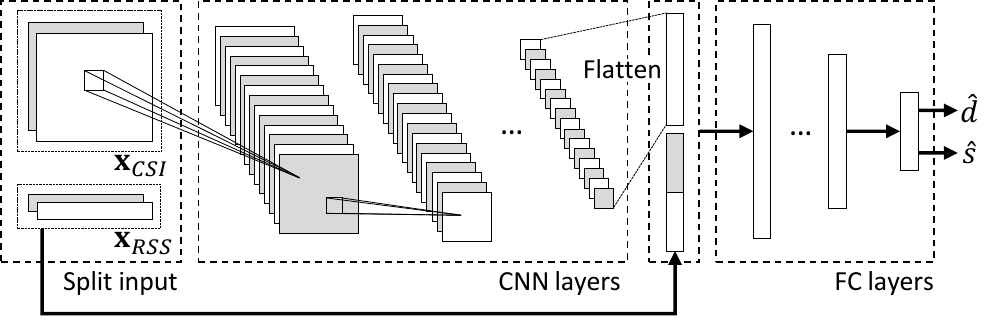}
    \caption{Ranging with convolutional neural networks.}
    \label{fig_CNN}
\end{figure}

Fig.~\ref{fig_CNN} illustrates the proposed ranging architecture.
The CNN layers extract features from the two channel CSI input image.
Because the number of collectible beacon frames from an AP is small in practice (e.g., $B\leq 10$), we use a $B\times 4$ sized kernel for the first convolution layer.
The output of the first convolution layer becomes a one dimensional vector, and we apply three more convolution layers with $1 \times 4$ sized kernels.
In addition, 64 filters are used for all the convolution layers, and the $1\times 2$ max-pooling layers are applied to every two convolution layers.
The output of the CNN layers is concatenated with the two RSS input vectors after the flatten operation.
Finally, three fully-connected (FC) layers, each with 256 hidden nodes, are applied to the concatenated layer to produce the outputs.

The distance estimate and its standard deviation are obtained from the activation of the last FC layer as follows:
\begin{align}
    \hat{d} =& \bar{d} \sigma(\mathbf W_d \tilde{\mathcal X} + b_d),\nonumber\\
    \hat{s} =& \bar{s} \sigma(\mathbf W_s \tilde{\mathcal X} + b_s),
\end{align}
where $\tilde{\mathcal X}\in \mathbb{R}^{256 \times 1}$ represents the activation vector of the last hidden layer corresponding to the input layer $\mathcal X$.
Matrices $\mathbf W_d$ and $\mathbf W_s \in \mathbb{R}^{256 \times 1}$ represent the weights between the last hidden layer and each output, and scalars $b_d$ and $b_s$ represent biases.
Throughout the NN architecture, the rectified linear unit (ReLU) is used for activation of every layer except for the output layer.
For output activation, the sigmoid function $\sigma(\cdot)$ is applied to specify the upper bounds of the distance and standard deviation outputs, denoted by $\bar{d}$ and $\bar{s}$, respectively.

\subsection{AP Offset Compensation}

The ranging module introduced in the previous subsection can be used to obtain ranging results for every AP.
However, each AP may use different transmission power, and the surrounding environment of each AP may differ. 
To address this, we introduce another trainable parameters that refer to the RSS offset of each AP.
We generate a vector of offsets where each element is associated one to one with each AP and prepare the input layer by adding the associated offset to every element in the RSS input.
In other words, the input layer for computing ranging results from the $n$-th selected AP at time step $k$ is expressed by
\begin{equation}
    \mathcal X_n^{(k)} = [\mathcal X_{CSI, n}^{(k)}, \mathcal X_{RSS, n}^{(k)} + o_n^{(k)}],
\end{equation}
where $\mathcal X_{CSI, n}^{(k)}$ and $\mathcal X_{RSS, n}^{(k)}$ represent the CSI and RSS inputs for the AP, respectively, and $o_n^{(k)}$ is the associated offset to the AP.
By feeding this input layer into the proposed ranging module, we can obtain the ranging results as $\mathcal R(\mathcal X_n^{(k)}; \Theta) = [\hat{d}_n^{(k)}, \hat{s}_n^{(k)}]^T$.
The offset of each AP is optimized during the training phase as errors propagate backward to the input layer.

\subsection{Positioning with Wi-Fi Ranging Results}

The ranging results from nearby APs are used to obtain the position of the device.
In this work, we apply an extended Kalman filter (EKF) that estimates unknown states using a series of measurements over time~\cite{8839041}.
The unknown state is assumed as the position of the device, and the EKF procedure is summarized as follows:

\emph{1) Initialization}: We initialize the estimate of the device's position as the center of $N$ nearby APs as
\begin{equation} \label{4c1}
    \hat{\mathbf z}^{(0)} = \frac{1}{N}\sum_{n=1}^{N} \mathbf z_n^{(0)},
\end{equation}
and the covariance matrix of the initial position estimate as
\begin{equation} \label{4c2}
    \mathbf P^{(0)} = \mbox{diag}\left(s_x^2, s_y^2\right),
\end{equation}
where $s_x, s_y$ represent the standard deviation of the initial x- and y-coordinate estimates.

\emph{2) State Prediction}: At each time step, the EKF predicts the current state based on the previous state using the state transition model given by
\begin{equation} \label{4c3}
    \mathbf z^{(k)} = \mathbf z^{(k-1)} + v\Delta t \mathbf u(\Phi),~k\geq 1,
\end{equation}
where $v$ represents the moving speed of the device that can be assumed as a constant, and $\Delta t$ is the time between two time steps.
Without external information, we simply express the direction of the device using a uniform random variable $\Phi$ that realizes a value from the interval $[0, 2\pi]$.
Using the state transition model, the predicted state and its covariance matrix at time step $k$ are obtained as
\begin{gather} \label{4c4}
    \hat{\mathbf z}^{(k|k-1)} = \hat{\mathbf z}^{(k-1)},\nonumber\\
    \mathbf P^{(k|k-1)} = \mathbf P^{(k-1)} + \mathbf Q^{(k)},
\end{gather}
where $\mathbf Q^{(k)} = E[(v\Delta t \mathbf u(\Phi))(v\Delta t \mathbf u(\Phi))^T] = \frac{1}{2}(v\Delta t)^2\mathbf I_{2}$.

\emph{3) State Update}: The predicted state is corrected with ranging results.
The measurement model is expressed as
\begin{equation} \label{4c5}
    \mathbf d^{(k)} = \mathbf h^{(k)}(\mathbf z) + \boldsymbol{\omega}^{(k)}
    = 
    \begin{bmatrix}
    \lVert \mathbf z - \mathbf z_1^{(k)}\rVert \\
    \vdots\\
    \lVert \mathbf z - \mathbf z_N^{(k)}\rVert
    \end{bmatrix} + \boldsymbol{\omega}^{(k)},
\end{equation}
where $\mathbf d^{(k)}=[\hat{d}_1^{(k)}, ..., \hat{d}_N^{(k)}]^T$ is a vector of distance estimates from $N$ nearby APs at time step $k$.
In addition, $\boldsymbol{\omega}^{(k)} = [\omega_1^{(k)}, ..., \omega_N^{(k)}]^T$ indicates a vector of ranging errors whose covariance matrix is computed with the second outputs of the ranging module as
\begin{equation} \label{4c6}
    \boldsymbol{\Lambda}^{(k)} = E\left[\boldsymbol{\omega}^{(k)}(\boldsymbol{\omega}^{(k)})^T\right] = \mbox{diag}\left((\hat{s}_1^{(k)})^2, ..., (\hat{s}_N^{(k)})^2\right).
\end{equation}
The innovation of the EKF is computed as
\begin{equation}
    \mathbf e^{(k)} = \mathbf d^{(k)} - \mathbf h^{(k)}(\hat{\mathbf z}^{(k|k-1)}),
\end{equation}
and the covariance matrix of the innovation is given by
\begin{equation}
    \mathbf S^{(k)} = \mathbf H^{(k)} \mathbf P^{(k|k-1)} (\mathbf H^{(k)})^T + \boldsymbol{\Lambda}^{(k)}.
\end{equation}
Here, $\mathbf H^{(k)} \in \mathbb{R}^{N\times 2}$ represents the Jacobian matrix of $\mathbf h^{(k)}(\cdot)$, which is defined as
\begin{equation}
    \mathbf H^{(k)} \triangleq \evalat[\Big]{\frac{\partial \mathbf h^{(k)}(\mathbf z)}{\partial \mathbf z}}{\mathbf z = \hat{\mathbf z}^{(k|k-1)}}.
\end{equation}

The Kalman gain is computed as
\begin{equation}
    \mathbf G^{(k)} = \mathbf{P}^{(k|k-1)}(\mathbf H^{(k)})^T(\mathbf{S}^{(k)})^{-1},
\end{equation}
and the updated state and its covariance matrix are given by
\begin{gather}
    \hat{\mathbf z}^{(k)} = \hat{\mathbf z}^{(k|k-1)} + \mathbf{G}^{(k)}\mathbf{e}^{(k)},\nonumber\\
    \mathbf{P}^{(k)} = \left(\mathbf{I}_{2} - \mathbf{G}^{(k)} \mathbf{H}^{(k)}\right)\mathbf{P}^{(k|k-1)},
\end{gather}
respectively. 
Note that $\hat{\mathbf z}^{(k)}$ is the estimated position of the device at time step $k$, and the shape of the estimated trajectory widely varies depending on the set of parameters in the ranging module $\Theta$.

\section{Learning Technique}

\begin{figure}
    \centering
    \includegraphics[width=0.45\textwidth]{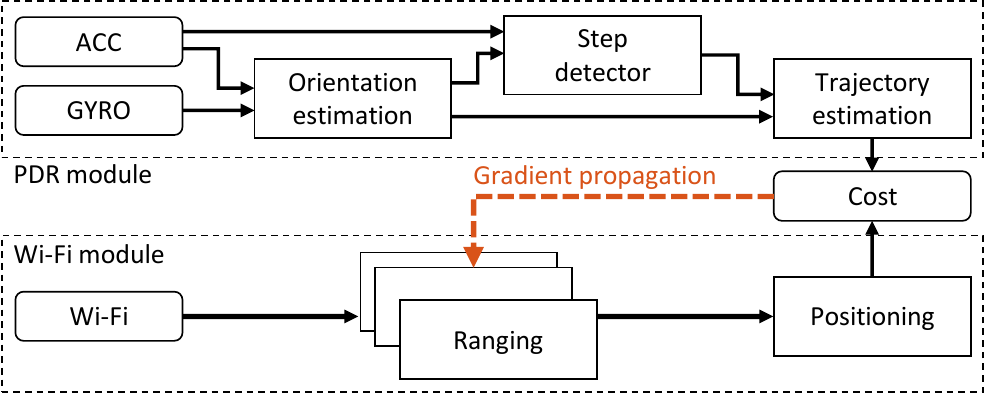}
    \caption{Overview of the proposed sensor-aided learning technique.}
    \label{fig_overview}
\end{figure}

Fig.~\ref{fig_overview} provides an overview of the sensor-aided learning technique.
For the training, the trajectory of the device is separately estimated using Wi-Fi and PDR modules. 
Then a cost function that compares the shape of the two trajectories is defined.
Because the sampling rate of sensors is much faster than the frequency of the Wi-Fi ranging procedure, we first synchronize the two trajectories by defining $\mathbf p^{(k)} \triangleq  \mathbf p(t_k)$ as the output of the PDR module at time $t_k$, which indicates the time at which the $k$-th Wi-Fi ranging procedure is performed.

In this section, we design a cost function of a single training dataset collected from time step 1 to $K$.
In case that multiple dataset are available, the exactly same process is applied to each dataset and the overall cost is obtained by simply tacking the average of all the costs.
For ease of exposition, we define $\mathcal{Z} \triangleq \{\hat{\mathbf z}^{(k)}\}_{k=1}^{K}$ as the sequence of estimated position of the device using the Wi-Fi module and $\mathcal{P} \triangleq \{\mathbf p^{(k)}\}_{k=1}^{K}$ as the sequence of the PDR output. In addition, all summation operations used in this section represent the summation from $k=1$ to $K$.

\begin{figure}
    \centering
    \includegraphics[width=0.40\textwidth]{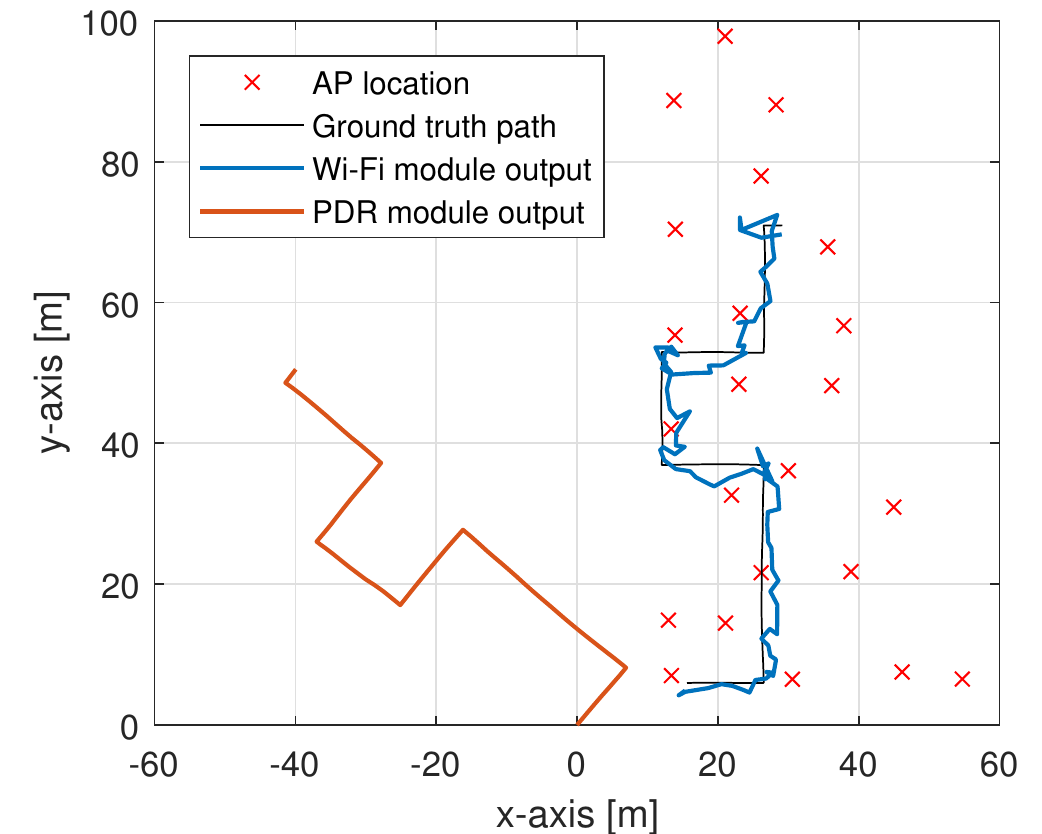}
    \caption{Example of trajectory estimation results using the Wi-Fi and PDR modules.}
    \label{fig_sen_example}
\end{figure}

Fig.~\ref{fig_sen_example} illustrates the trajectory estimation results.
Because the ranging module produces errors, the estimated trajectory using the Wi-Fi module fluctuates widely.
Nevertheless, the estimated trajectory follows the ground truth path shown in the figure as the positioning module exploits the true location of APs.
On the other hands, the PDR module produces a smoother trajectory because built-in sensors are less affected by the external environment, and the shape of the trajectory is similar to that of the ground truth path.
However, the estimated trajectory starts at an arbitrary position, and the direction of the movement is not aligned with the ground truth path.

To address this issue, we first transform the PDR output with a rotation angle and an offset as follows:
\begin{equation} \label{5A1}
    \tilde{\mathbf p}^{(k)} = \mathbf R(\varphi)\mathbf p^{(k)} + \boldsymbol{\Omega},~1\leq k \leq K,
\end{equation}
where $\boldsymbol{\Omega} \in \mathbb{R}^{2\times 1}$ indicates the x- and y-coordinate offsets, and $\varphi$ is the rotation angle.
In addition, $\mathbf R(\varphi) \in \mathbb{R}^{2\times 2}$ represents the rotation matrix on the x-y plane, which is defined as
\begin{equation} \label{5A2}
    \mathbf R(\varphi) \triangleq
    \begin{bmatrix} \cos\varphi & -\sin\varphi \\ \sin\varphi & \cos\varphi
    \end{bmatrix}
    = \cos\varphi \mathbf I_{2} + \sin\varphi \tilde{\mathbf I}_{2},
\end{equation}
with $\tilde{\mathbf I}_{2} \in \mathbb{R}^{2 \times 2}$ given by
\begin{equation} \label{5A3}
    \tilde{\mathbf I}_{2} \triangleq \begin{bmatrix} 0 & -1 \\ 1 & 0 \end{bmatrix}.
\end{equation}
An optimal transformation can be obtained to minimize the sum squared error between the transformed PDR output and estimated trajectory using the Wi-Fi module, which is given by
\begin{equation} \label{5A4}
    J\left(\mathcal Z, \mathcal P; \varphi, \boldsymbol{\Omega}\right) = \sum_{k} \lVert \tilde{\mathbf p}^{(k)} - \hat{\mathbf{z}}^{(k)} \rVert^2.
\end{equation}
We use the following lemma to obtain an optimal transformation.

\emph {Lemma~1}: An optimal rotation angle and offset that minimize the cost function in equation (\ref{5A4}) are derived as
\begin{gather} 
    {\varphi}^*= \pi + \arctan\frac{\Gamma}{\tilde{\Gamma}},\nonumber\\
    {\boldsymbol{\Omega}}^* = \frac{\sum_k \hat{\mathbf{z}}^{(k)} - \mathbf{R}(\varphi^*)\sum_k \mathbf p^{(k)}}{K},
    \label{5A5}
\end{gather}
respectively, where $\Gamma$ and $\tilde{\Gamma}$ are related to $\mathcal Z$ and $\mathcal P$ as follows:
\begin{gather} 
    \Gamma = \frac{(\sum_k \hat{\mathbf z}^{(k)})^T(\sum_k \mathbf p^{(k)})}{K} - \sum_k (\hat{\mathbf z}^{(k)})^T \mathbf p^{(k)},\nonumber\\
    \tilde{\Gamma} = \frac{(\sum_k \hat{\mathbf z}^{(k)})^T\tilde{\mathbf I}_2(\sum_k \mathbf p^{(k)})}{K} - \sum_k (\hat{\mathbf z}^{(k)})^T \tilde{\mathbf I}_2 \mathbf p^{(k)}.
    \label{5A6}
\end{gather}
The error after the transformation is computed as
\begin{align}\label{5A7}
    J(\mathcal Z, \mathcal P; \varphi^*, \boldsymbol{\Omega}^*) = \sum_k \lVert \hat{\mathbf z}^{(k)}\rVert^2 + \sum_k \lVert \mathbf p^{(k)}\rVert^2&\nonumber\\
    + \frac{\lVert\sum_k \hat{\mathbf z}^{(k)} \rVert + \lVert\sum_k \mathbf p^{(k)} \rVert}{K} - 2\sqrt{\Gamma^2 + \tilde{\Gamma}^2}&.
\end{align}

\begin{proof}
See Appendix~A.
\end{proof}

This lemma explains that if two estimated trajectories are given, one can be transformed close to the other to compute the cost that measures the similarity of the shape of the two trajectories.
Using this cost, we can train the ranging module so that the shape of the estimated trajectory using the Wi-Fi module becomes similar to the shape of the PDR output, which is almost the same as the shape of the ground truth path.
Based on this observation, we can define a cost function that utilizes sensor information as
\begin{equation} \label{5A8}
    {J}^{sen}(\mathcal Z, \mathcal P) = J(\mathcal Z, \mathcal P; \varphi^*, \boldsymbol{\Omega}^*),
\end{equation}
which is computed using equation~(\ref{5A7}).

In addition to the above cost function, we can also reuse the geometric cost function introduced in~\cite{8839041}.
This cost function depends only on the estimated trajectory using the Wi-Fi module as
\begin{equation}
    J^{geo}(\mathcal Z) = \sum_{k=1}^K\sum_{n=1}^N\left(\lVert\hat{\mathbf z}^{(k)} - \mathbf z_n^{(k)}\rVert - \hat{d}_n^{(k)}\right)^2.
\end{equation}
By combining the two cost functions, we can obtain a unified cost function for training as
\begin{equation}
    J(\mathcal Z, \mathcal P) = \mu_1 J^{sen}(\mathcal Z, \mathcal P) + \mu_2J^{geo}(\mathcal Z),
\end{equation}
where $\mu_1$ and $\mu_2$ are non-negative constants that balance the two cost functions.
Because the unified cost function depends on $\mathcal Z$ and $\mathcal P$, where $\mathcal P$ is fixed as the PDR output, the gradient of the unified cost function with respect to every element in $\mathcal Z$ can be easily computed, and these gradients propagate to the ranging module to optimize every trainable parameter.

\section{Experimental Results}

\begin{figure}
    \centering
    \subfloat[]{\includegraphics[width=0.24\textwidth]{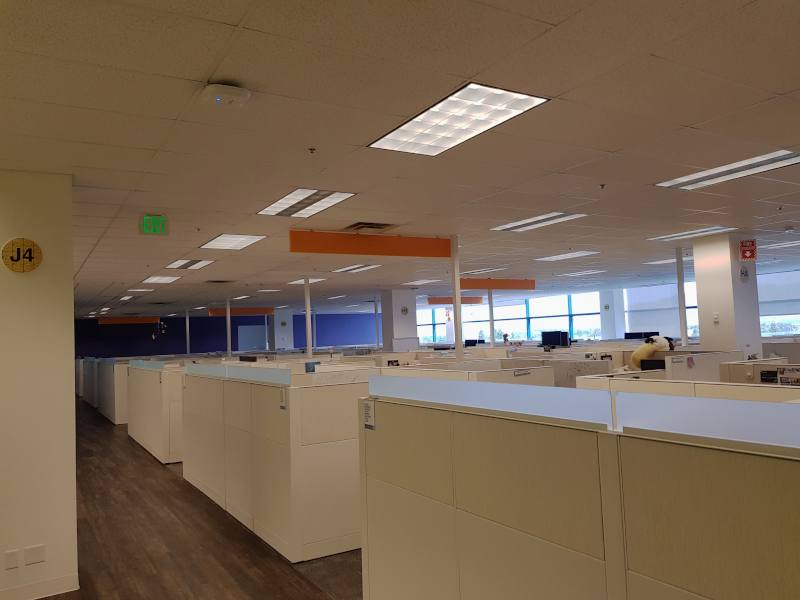}}\hfil\hfil
    \subfloat[]{\includegraphics[width=0.24\textwidth]{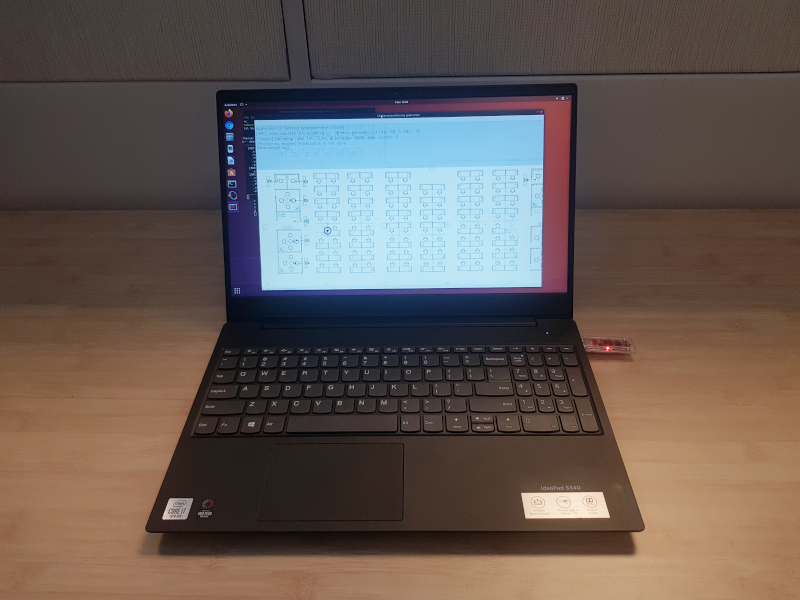}}
    \caption{Experimental site and device. (a) Office environment with multiple Wi-Fi APs installed on the ceiling and (b) laptop equipped with Intel AX200 Wi-Fi chipset and an external USB sensor stick.}
    \label{fig_exp}
\end{figure}

\subsection{Experiment Setup}

We performed measurement campaigns in a practical indoor office environment with 59 Wi-Fi APs installed on the ceiling.
A laptop running Ubuntu 18.04 operating system was used for the experiments.
This laptop is equipped with an Intel Wi-Fi 6 AX200 chipset that supports all IEEE 802.11 standards up to 11ax.
We also implemented a real-time positioning application to collect training data and to demonstrate the proposed method.
Fig.~\ref{fig_exp} shows pictures of the experiment site and the device with the real-time application running on the screen.

We used the new CSI tool mentioned in Introduction.
Because this tool can collect the CSI of any incoming packets, the easiest way to capture the CSI of the beacon frame is to perform the channel scanning procedure by executing the \emph{iw dev scan} command on the terminal (or \emph{iw dev scan freq} to specify the list of channels to be scanned).
However, for more flexible control, we configured the Wi-Fi interface to the monitor mode for this experiment to capture all packet transmissions on a specific Wi-Fi channel monitored by the Wi-Fi chipset.

The existing APs use three non-overlapping Wi-Fi channels for the 2.4~GHz frequency band, which are 1, 6, and 11. 
Therefore, we allocated 300~ms to each channel to collect beacon frames transmitted at that channel; thus, a single ranging and positioning procedure took less than 1 s.
Furthermore, each AP used in this experiment site transmits 4 different service set identifiers (SSIDs) on the 2.4~GHz frequency band, and each SSID is broadcast with a 100~ms beacon interval.
Therefore, it was possible to receive multiple beacon frames from nearby APs during a single ranging procedure.
The positioning performance was evaluated with various choices of the number of beacon frames used in the ranging module (e.g., $B=1, 2, 4,$ and  $8$).
However, the performance was almost the same regardless of $B$.
Thus, we assume that $B=4$ throughout the experiment.
In addition, we use up to $N=5$ nearby APs in the positioning module.

Because built-in sensors on Intel mobile processors, called integrated sensor hub (ISH), have not yet been officially supported for the Ubuntu environment, we used an external USB sensor stick from Bosch Sensortech. 
Using a Python library provided in~\cite{bno055}, it was possible to collect accelerometer and gyroscope readings with a sampling rate of 100~Hz.
The coefficient for the step length is given by $\alpha = 0.55$.

\subsection{Training Phase}

\begin{table}
    \renewcommand{\arraystretch}{1.3}
    \caption{Wi-Fi Ranging Scenarios}
    \centering
    \begin{tabular}{ccc}
        \bf Ranging model & \bf Source & \bf Training Data\\
        \hline
        Path loss~\cite{Wang2003AnIW}  & RSS & Calibration data \\
        Polynomial~\cite{5425237}  & RSS & Calibration data \\
        CUPID~\cite{10.1145/2462456.2464463}  & RSS, CSI & Calibration data \\
        FC (unsupervised)~\cite{8839041}    & RSS & Unlabeled data\\
        FC (sensor-aided)    & RSS &  Unlabeled data (w/ sensor)\\
        CNN (sensor-aided)  & RSS, CSI & Unlabeled data (w/ sensor) \\
        \hline
    \end{tabular}
    \label{tab_ranging_scenario}
\end{table}

\begin{figure}
    \centering
    \includegraphics[width=0.40\textwidth]{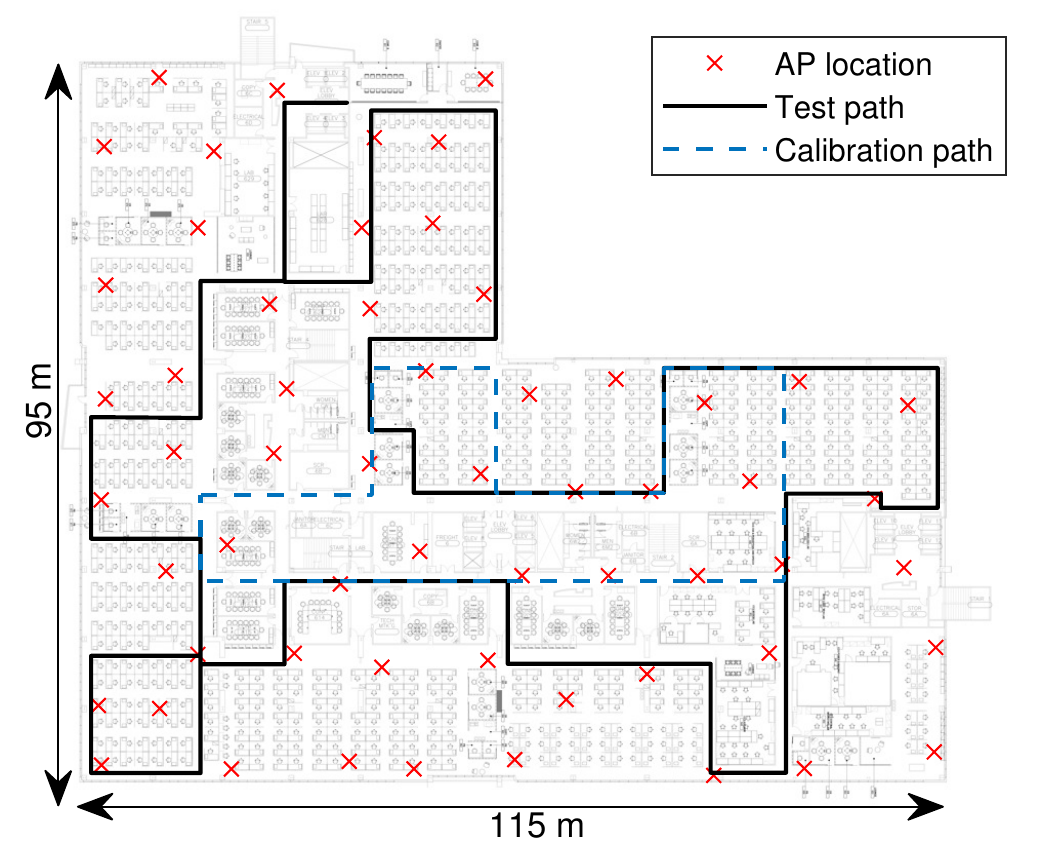}
    \caption{Floor plan of the experiment site.}
    \label{fig_map}
\end{figure}

For performance comparisons, we considered various ranging scenarios, as summarized in Table~\ref{tab_ranging_scenario}. 
First, the path loss-based ranging model estimates the distance from an AP as~\cite{Wang2003AnIW}
\begin{equation} \label{6b1}
    \hat{d}^{PL} = d_0 10^{\frac{RSS(d_0) - RSS}{10 \eta}},
\end{equation}
where $RSS$ represents the average of all RSS measurements using the two antennas, $RSS(d_0)$ is the RSS at a reference distance $d_0 = 1$~m, and $\eta$ is the path loss exponent.
In addition, the distance from an AP can be estimated using a quadratic polynomial as~\cite{5425237} 
\begin{equation}
    \hat{d}^{poly} = g_2 RSS^2 + g_1 RSS + g_0,
\end{equation}
where $g_2, g_1$, and $g_0$ represent the coefficients of the polynomial, which should be chosen appropriately.

\begin{figure}
    \centering
    \subfloat[]{\includegraphics[width=0.24\textwidth]{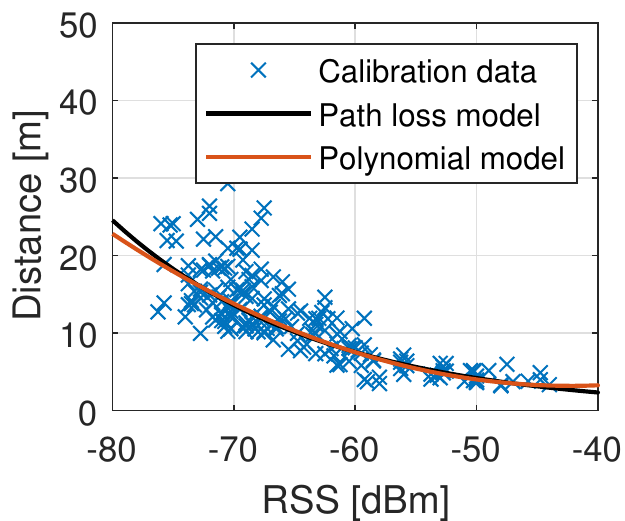}}\hfil    \subfloat[]{\includegraphics[width=0.24\textwidth]{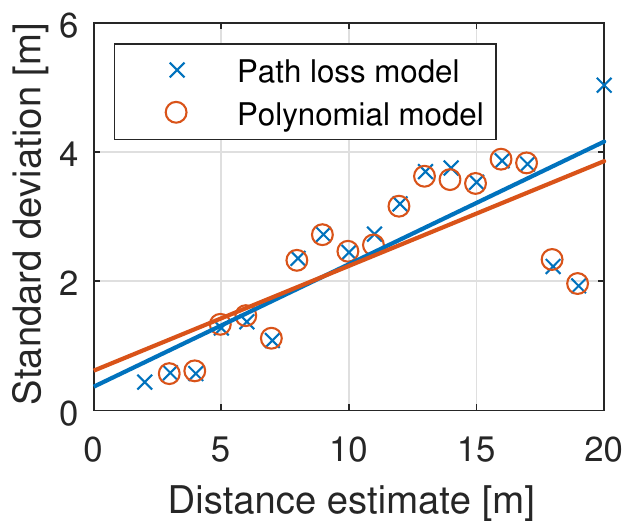}}
    \caption{Calibration of path loss and polynomial parameters: (a) RSS versus distance, and (b) distance estimate versus standard deviation.}
    \label{fig_pathloss}
\end{figure}

To optimize the parameters in the path loss and polynomial-based ranging models, we collected ground truth training data by following the calibration path presented in Fig.~\ref{fig_map}.
The collected data are the true x- and y-coordinates of the device and the received beacon frames at each position.
Fig.~\ref{fig_pathloss}(a) depicts the relationship between the measured RSS and the ground truth distance. 
The parameters in the path loss and polynomial models were selected to minimize the normalized mean squared error (NMSE) between the distance estimate $\hat{d}$ and the true distance $d^*$, which is defined as $NMSE = E[((\hat{d} - d^*)/d^*)^2]$.
The selected parameters for the path loss model are given by $RSS(d_0) = -25.8$~dBm and $\eta=3.9$, and those for the polynomial model are given by $g_2 = 0.0138$, $g_1 = 1.1642$, and $g_0 = 27.7688$.
The path loss and polynomial curves with the selected parameters are presented as solid lines in Fig.~\ref{fig_pathloss}(a).

\begin{figure*}
    \centering
    \subfloat[]{\includegraphics[width=0.30\textwidth]{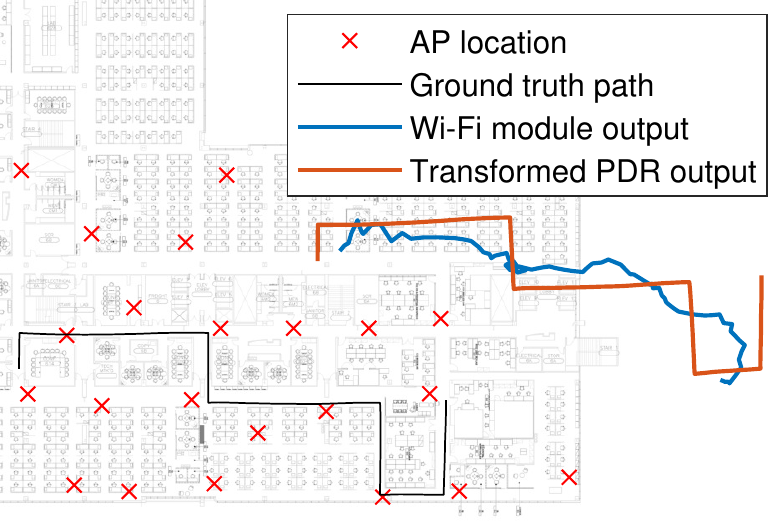}}\hfil\hfil
    \subfloat[]{\includegraphics[width=0.30\textwidth]{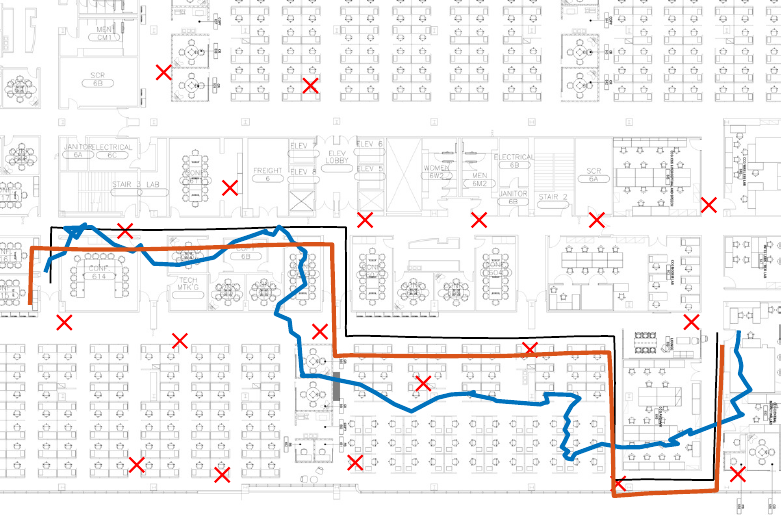}}\hfil\hfil
    \subfloat[]{\includegraphics[width=0.30\textwidth]{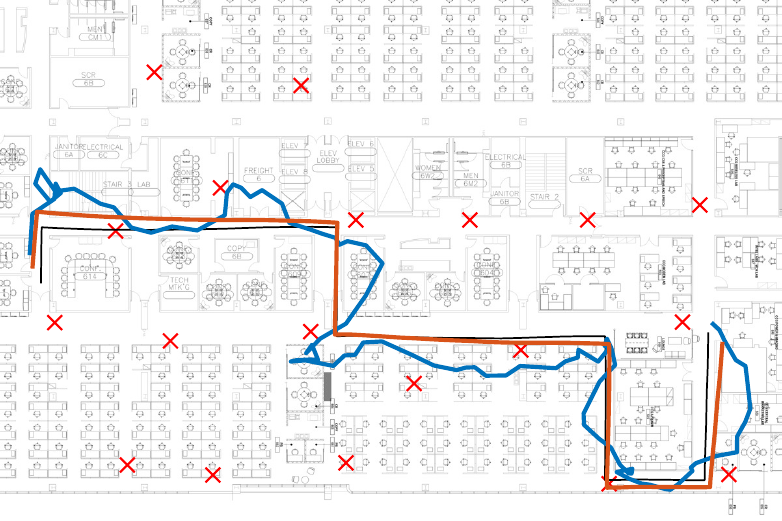}}\\
    \subfloat[]{\includegraphics[width=0.30\textwidth]{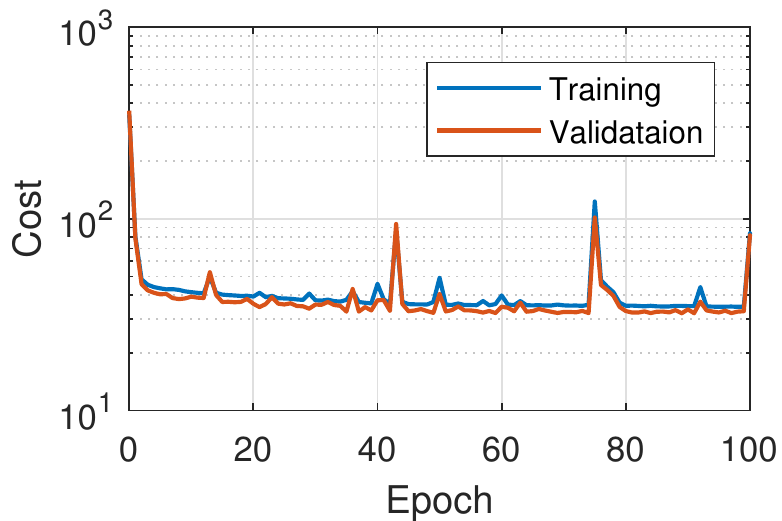}}\hfil\hfil
    \subfloat[]{\includegraphics[width=0.30\textwidth]{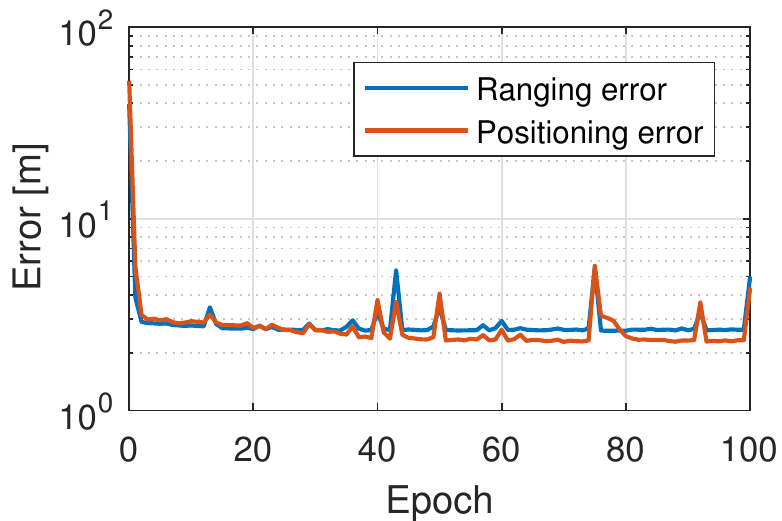}}\hfil\hfil
    \subfloat[]{\includegraphics[width=0.30\textwidth]{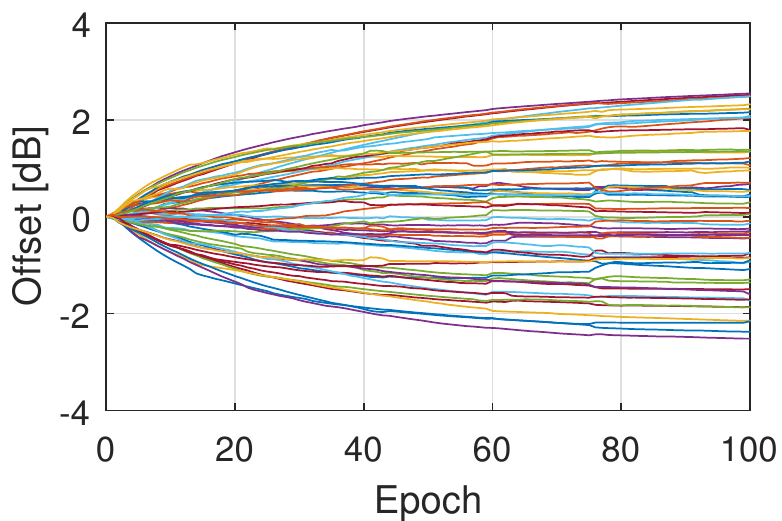}}
    \caption{Training details of the sensor-aided learning: (a)-(c)~Estimated trajectories at epoch 0, 5, and 50, respectively, (d)~training and validation cost, (e)~ranging and positioning error with respect to the test data, and (f)~trained offset of each AP.}
    \label{fig_sen_result}
\end{figure*}

In addition, the standard deviation of each distance estimate is empirically obtained by tacking the standard deviation of the ranging errors with respect to all data whose distance estimates are less than 1~m from the target distance estimate.
Fig.~\ref{fig_pathloss}(b) illustrates the relationship between the estimated distance and the empirically obtained standard deviation.
A linear regression line was used to make a model $\hat{s}^{PL} = 0.1897\hat{d}^{PL} + 0.3672$ for the path loss and $\hat{s}^{poly} = 0.1622\hat{d}^{poly} + 0.6156$ for the polynomial-based ranging scenarios.
In the same way, we optimized the parameters for the CUPID model that exploits both RSS and CSI for the ranging procedure~\cite{10.1145/2462456.2464463}.
To this end, the energy of the direct path (EDP) was extracted from the CSI, and different path loss exponent was selected depending on the ratio between EDP and RSS to estimate the distance using the path loss model.
The standard deviation for CUPID was modeled in the same way as with the previous models.

In addition to the model-based ranging scenarios, we also evaluated the performance with NN-based ranging scenarios.
To produce ranging results using RSS only, FC layers were deployed that consist of two hidden layers, each with 128 hidden nodes. 
The size of the input layer is $2B$ in this case.
We trained the FC layers using unsupervised learning~\cite{8839041} and sensor-aided learning methods to verify the effectiveness of the proposed learning technique.
Furthermore, we also verified the performance of the CNN-based ranging scenario using the CSI of beacon frames.
The maximum distance estimate and standard deviation were assumed to be $\bar{d} = 100$~m and $\bar{s}=10$~m for all NN-based ranging scenarios.

The calibration data were not used for training NN-based ranging modules; rather unlabeled training data were collected by randomly moving around the experiment site to imitate that users use a positioning application in practice.
Through several measurement campaigns, the training data were collected for 3600 time steps, and they were partitioned into multiple dataset, each with 100 time steps.
In the experiment, 70\% of the dataset were used for training and the remaining 30\% for validation. Adam optimizer with a learning rate of 0.001 was used for training.
We assumed that $(\mu_1, \mu_2) = (1, 1)$ for sensor-aided learning scenarios and $(\mu_1, \mu_2)=(0, 1)$ for unsupervised learning scenarios.
The test data were obtained by following the test path shown in Fig.~\ref{fig_map} at a constant speed.
We measured the time at each vertex of the test path; thus, the true coordinates of the device could be obtained using interpolation.

\begin{figure*}
    \centering
    \subfloat[]{\includegraphics[width=0.30\textwidth]{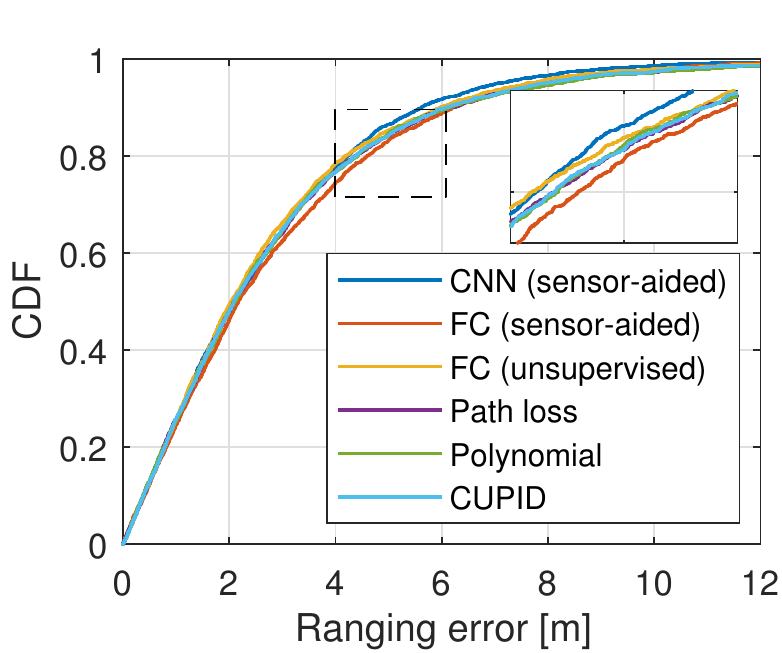}}\hfil\hfil
    \subfloat[]{\includegraphics[width=0.30\textwidth]{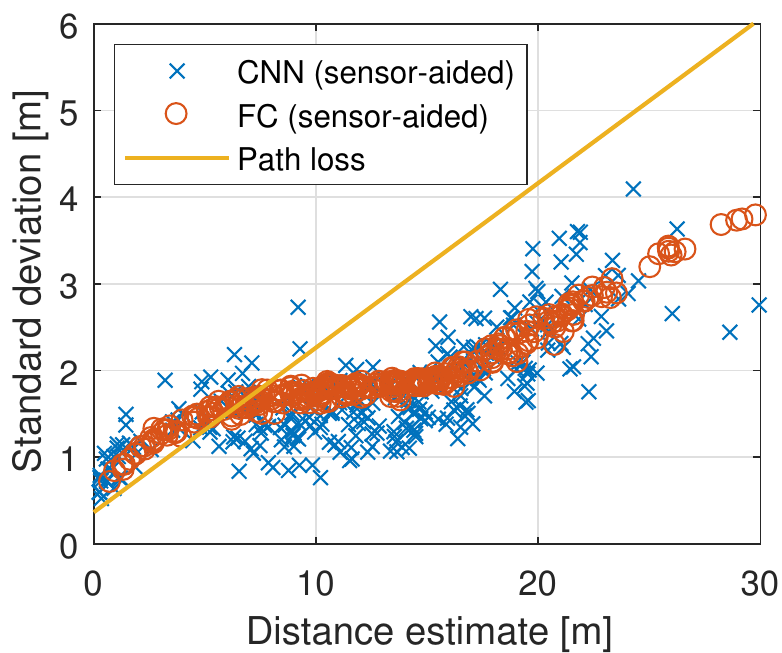}}\hfil\hfil
    \subfloat[]{\includegraphics[width=0.30\textwidth]{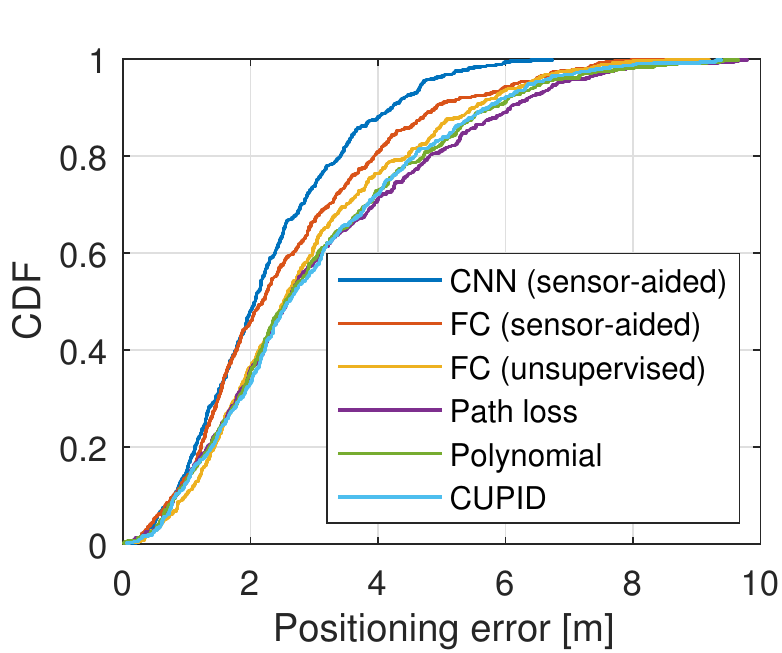}}
    \caption{Training results: (a) CDF of ranging error, (b) relationship between distance estimate and standard deviation, and (c) CDF of positioning error.}
    \label{fig_tr_summary}
\end{figure*}

Fig.~\ref{fig_sen_result} illustrates the details of the sensor-aided learning technique.
The parameters in the ranging module are randomly initialized in the beginning. 
As a result, the estimated trajectory using the Wi-Fi module produces an incorrect trajectory at epoch 0, as shown in Fig.~\ref{fig_sen_result}(a). Accordingly, the PDR output is transformed close to the Wi-Fi trajectory and the cost is computed.
After a few training epochs, the estimated trajectory using the Wi-Fi module closely approaches the test path because the ranging module produces accurate ranging results, and the transformed PDR output overlaps the test path, as shown in Fig.~\ref{fig_sen_result}(b) and (c).
Therefore, the cost computed using equation~(\ref{5A8}) can be equivalently considered as the error between the estimated trajectory using the Wi-Fi module and the test path.
Fig.~\ref{fig_sen_result}(d) illustrates that the costs with respect to training and validation data decrease with epoch.
As a result, the ranging and positioning errors with respect to the test data also decrease with epoch, as shown in Fig.~\ref{fig_sen_example}(e). In addition, the offsets of all 59 APs are optimized during the training phase, as shown in Fig.~\ref{fig_sen_example}(f).

Fig.~\ref{fig_tr_summary}(a) depicts the cumulative density function (CDF) of the ranging error.
Because every scenario primarily relies on the RSS for the ranging, there was no significant difference among the ranging results.
For many scenarios, CNN-based ranging with beacon CSI produces the best CDF curve.
Fig.~\ref{fig_tr_summary}(b) shows more interesting results. This figure shows the relationship between the distance estimate and the standard deviation for selected ranging scenarios. 
The relationship between the two outputs for the path loss-based ranging scenario is presented as a straight line as we modeled it using a linear regression line.
However, the sensor-aided learning technique optimizes trainable parameters so that the shape of the estimated trajectory using ranging results becomes similar to that of the PDR output.
In this process, NN-based ranging modules autonomously learn the way to classify the current channel condition based on information provided in the input layer and produce various standard deviation outputs, even for the same distance estimate.
For instance, if information in the input layer is highly likely to be observed in an NLOS condition, the ranging module may produce a high standard deviation to make the positioning module less reliant on the current distance estimate.

\begin{table}
    \renewcommand{\arraystretch}{1.3}
    \caption{Ranging and Positioning Performance using Wi-Fi only}
    \centering
    \begin{tabular}{c|cccc}
        &\multirow{2}{*}{\bf Ranging model} & \bf MAE & \bf RMSE & \bf 90\%-tile\\
        & & \bf [m] & \bf [m] & \bf [m]\\
        \hline
        {\multirow{6}{*}{\rotatebox[origin=c]{90}{Ranging}}}&Path loss  & 3.019 &  4.175 & 6.746 \\
        &Polynomial & 3.011 &  4.203 & 6.578 \\
        &CUPID  & 2.971 &  4.111 & 6.659\\
        &FC (unsupervised) & 2.717  & 3.813 &  6.127\\
        &FC (sensor-aided)    & 2.874 & 3.882 & 6.206  \\
        &CNN (sensor-aided)  & 2.657 & 3.602 & 5.628 \\
        \hline
        {\multirow{6}{*}{\rotatebox[origin=c]{90}{Positioning}}}&Path loss  & 2.887 & 3.442 & 5.436 \\
        &Polynomial & 2.851 &  3.400 & 5.300 \\
        &CUPID  & 2.880 &  3.411 & 5.338 \\
        &FC (unsupervised)    & 2.967 &  3.468 & 5.436\\
        &FC (sensor-aided)    & 2.595 &  3.079 & 5.008  \\
        &CNN (sensor-aided)  & 2.300 &  2.626 & 3.903 \\
        \hline
    \end{tabular}
    \label{tab_wifi_result}
\end{table}

Fig.~\ref{fig_tr_summary}(b) shows that the range of standard deviation for the CNN-based ranging scenario is wider than that for the FC scenario. This means that the CNN layer identifies more diverse channel conditions from the CSI of beacon frames to produce outputs.
The results of diverse standard deviation estimation can be observed in the positioning performance.
Fig.~\ref{fig_tr_summary}(c) depicts the CDF of the positioning error.
Although all ranging scenarios yielded similar ranging performance, the NN-based ranging modules trained with the sensor-aided learning technique outperform the existing methods as they learned how to identify the current channel condition and produce a precise standard deviation depending on the channel condition.

\begin{figure}
    \centering
    \includegraphics[width=0.40\textwidth]{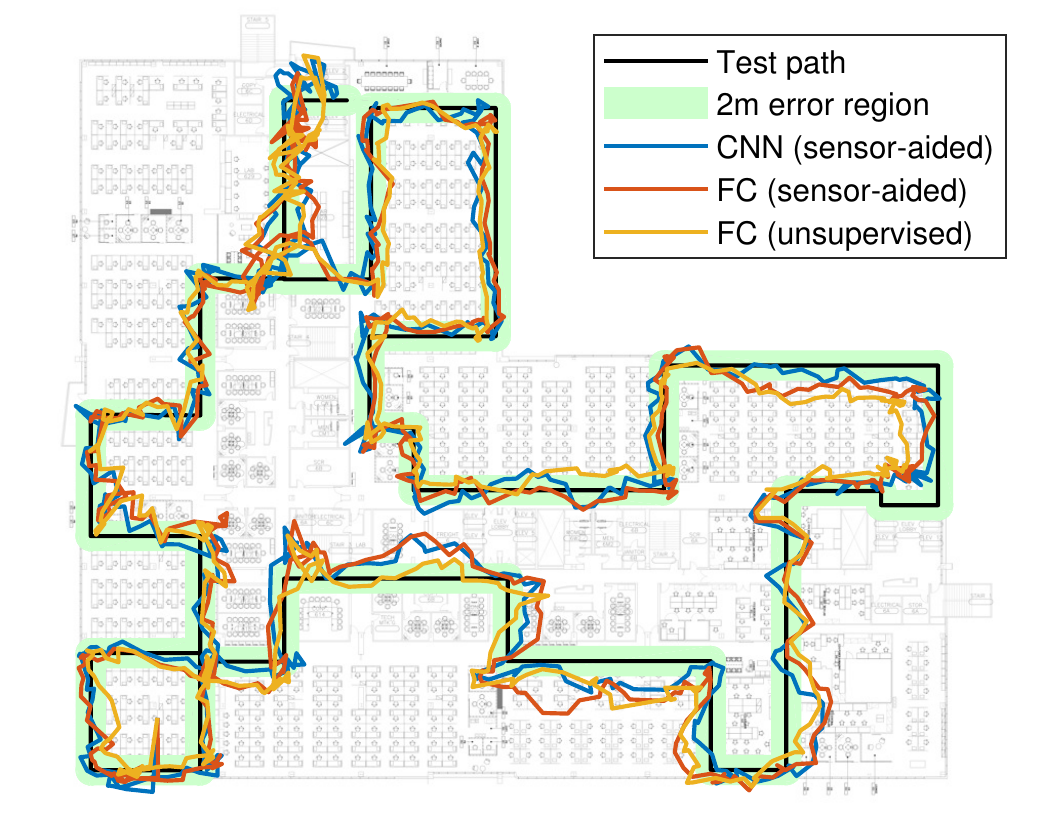}
    \caption{Estimated trajectory using Wi-Fi only.}
    \label{fig_tr_trj}
\end{figure}

The ranging and positioning performances of the scenarios are summarized in Table~\ref{tab_wifi_result}.
Note that the positioning results summarized in the table rely only on the Wi-Fi module.
The performance metrics are the mean absolute error (MAE), root mean squared error (RMSE), and 90-th percentile error. 
Finally, Fig.~\ref{fig_tr_trj} shows the estimated trajectory of the selected scenarios.
The green area represents the 2~m error region, indicating that any points in this area are less than 2~m apart from the closest test path.
The estimated trajectory for the CNN-based ranging scenario produces the best performance in terms of every performance metric.



\subsection{Online Phase}

Once training is completed, the Wi-Fi ranging module produces distance and standard deviation estimates for all nearby APs during the online phase.
In the case where the Wi-Fi ranging module is only involved in the positioning process, we can achieve exactly the same performance as that summarized in Table~\ref{tab_wifi_result}.
In this section, we discuss the positioning performance with Wi-Fi ranging and sensors together.

One issue is that the heading angle of the device, obtained using accelerometer and gyroscope readings, is reported relative to an arbitrary reference direction that is not aligned with the x- and y-axes of the GCS in general.
Therefore, we can include the unknown reference direction in the EKF state to estimate it with a series of measurements.
According to the experimental results in~\cite{9125898}, the EKF can estimate the correct reference direction using Wi-Fi ranging results if it is well-initialized.
To further improve the reliability of the initial estimation of the reference direction, we modify the EKF design by considering multiple candidates of reference directions simultaneously and selecting the best one depending on the ranging results.
The detailed process is summarized in Appendix~B.

\begin{figure}
    \centering
    \subfloat[]{\includegraphics[width=0.24\textwidth]{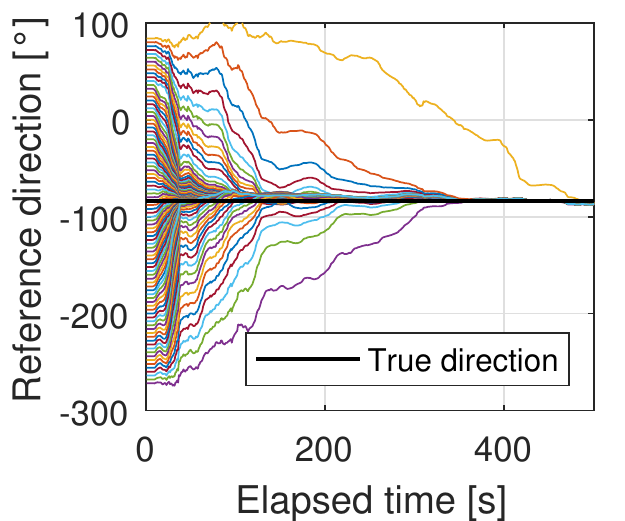}}\hfil\hfil    \subfloat[]{\includegraphics[width=0.24\textwidth]{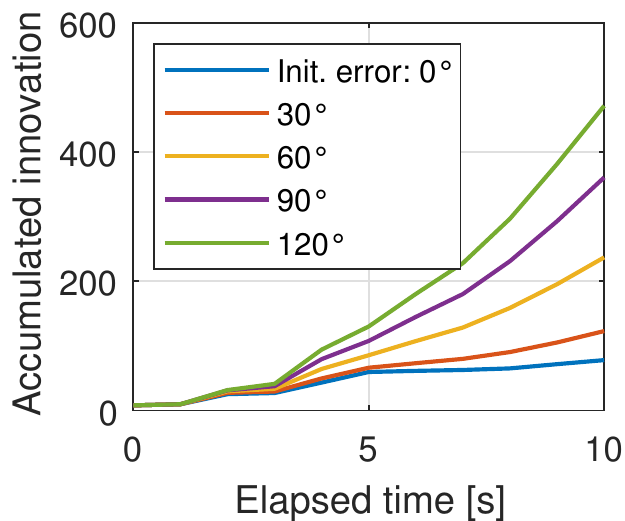}}
    \caption{Reference heading direction estimation with different initializations: (a) Reference direction and (b) accumulated innovation for 10 s.}
    \label{fig_online_ekf}
\end{figure}

In this experiment, 90 different candidates for the reference direction were generated with equally partitioned angles from 0$^{\circ}$ to 360$^{\circ}$.
Fig.~\ref{fig_online_ekf}(a) illustrates that different initializations of the reference directions converge to the true reference direction over time.
However, the reference directions initialized far from the true direction converge slowly compared with those initialized near the true direction.
As a result, the estimated trajectory with a wrong initial reference direction estimate produces an incorrectly estimated trajectory until the reference direction estimate approaches the true direction closely.
Fig.~\ref{fig_online_ekf}(b) shows the accumulated innovation of the EKF for selected candidates, where reference directions are initialized with different errors from the true reference direction.
The more errors in the initial reference direction estimate, the longer the trajectory estimation error, which produces a high innovation during the EKF process.
Based on this observation, we select the best reference heading direction candidate at each time step based on the innovation.
Furthermore, we can track the only candidate that is selected as the best at a sufficiently long time after initialization (e.g., 10 s) to reduce the computational complexity.

\begin{figure}
    \centering
    \includegraphics[width=0.40\textwidth]{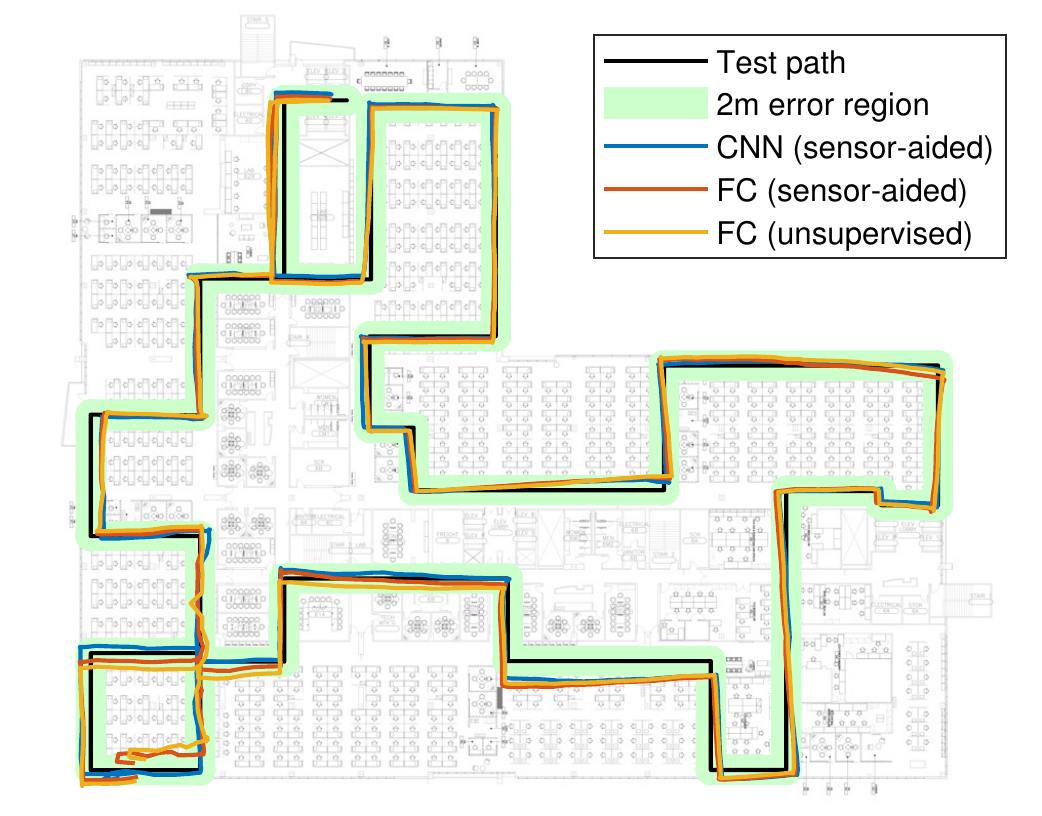}
    \caption{Estimated trajectory using Wi-Fi and sensors.}
    \label{fig_online_trj}
\end{figure}

\begin{figure}
    \centering
    \includegraphics[width=0.40\textwidth]{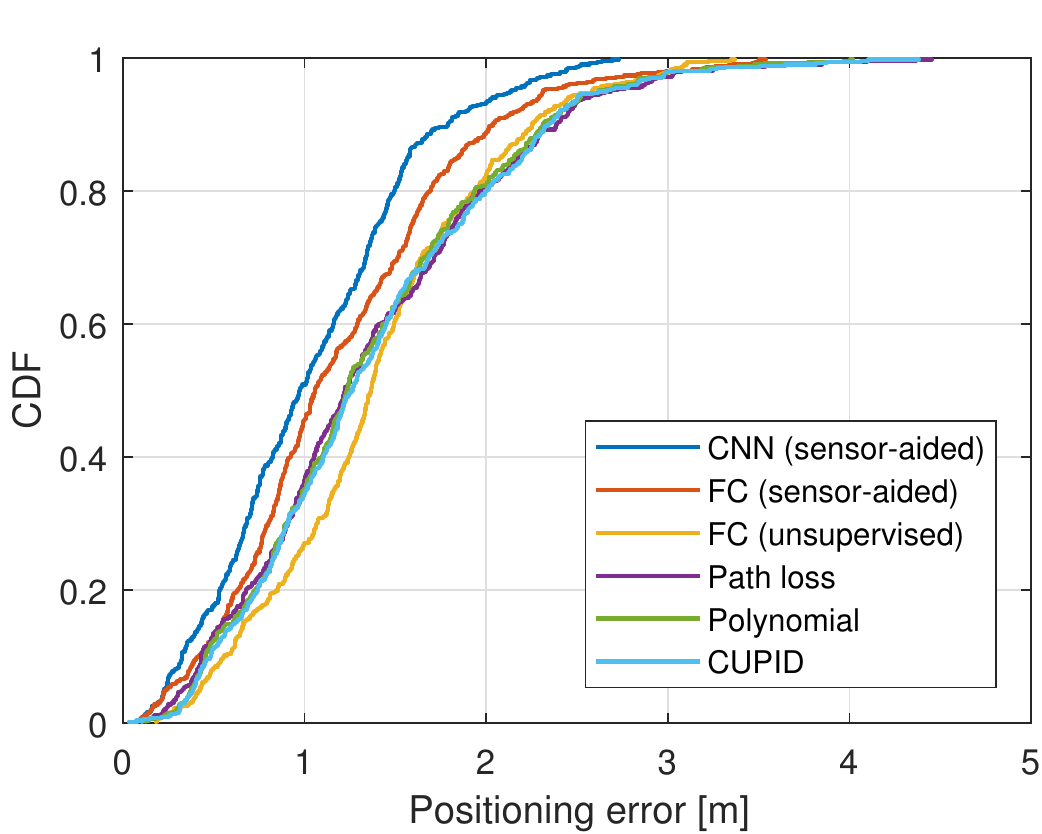}
    \caption{CDF of positioning error using Wi-Fi and sensors.}
    \label{fig_online_cdf}
\end{figure}

Fig.~\ref{fig_online_trj} illustrates the estimated trajectory using Wi-Fi ranging and the PDR technique for the selected scenarios.
Because PDR with estimated reference direction provides an accurate trajectory of the device, it is possible to obtain a smoother trajectory compared with the positioning results using Wi-Fi only.
Finally, Fig.~\ref{fig_online_cdf} depicts the CDF of the positioning error for every scenario, and Table~\ref{tab_online_summary} summarizes the results.
With the PDR technique, the positioning performance of every scenario was improved from the positioning results using the Wi-Fi module only. 
In particular, the CNN-based ranging scenario yielded the best performance for all metrics because it produced precise distance and standard deviation estimates.

\begin{table}
    \renewcommand{\arraystretch}{1.3}
    \caption{Positioning Performance with Wi-Fi Ranging and Sensors}
    \centering
    \begin{tabular}{cccc}
        \bf Ranging method & \bf MAE [m] & \bf RMSE [m] & \bf 90\%-tile [m]\\
        \hline
        Path loss  & 1.356 & 1.552 & 2.384  \\
        Polynomial  & 1.351 & 1.529 & 2.311 \\
        CUPID &  1.373 & 1.555 & 2.340   \\
        FC (unsupervised)    & 1.403 & 1.543 & 2.252 \\
        FC (sensor-aided)    & 1.192  & 1.362 & 2.034  \\
        CNN (sensor-aided)  &  1.038 & 1.180 & 1.787  \\
        \hline
    \end{tabular}
    \label{tab_online_summary}
\end{table}

\section{Conclusion}

In this paper, we studied an unsupervised learning technique to optimize a Wi-Fi ranging module using the sensor readings generated in a mobile device.
Because the PDR technique provides an accurate shape of the device trajectory, which is almost the same as the shape of the ground truth path, the output of the PDR module could be used as a reference in the training phase.
With the proposed cost function that measures the similarity between the estimated trajectory using the Wi-Fi module and PDR output, the ranging module autonomously learned how to identify the current channel condition and produce accurate ranging results accordingly.
In addition to RSS, the CSI of beacon frames was also collected using the new CSI tool with the latest Intel Wi-Fi chipset, and the benefit of using CSI for positioning was verified.
We believe that the CSI of the beacon frame can improve the positioning performance in more complicated indoor environments because it has more information about the current channel than RSS.
In addition, the proposed learning technique can significantly minimize human intervention in training data collection and site survey processes, and the positioning accuracy will improve as training data quickly accumulate as many users use the positioning application.

\appendices

\section{Proof of Lemma 1}

An offset that minimizes the cost function in equation~(\ref{5A4}) should satisfy the following relationship:
\begin{equation}
    \frac{\partial J(\mathcal Z, \mathcal P; \varphi, \boldsymbol{\Omega})}{\partial \boldsymbol{\Omega}} = \sum_{k} 2(\tilde{\mathbf p}^{(k)} - \hat{\mathbf z}^{(k)}) = [0, 0]^T.
\end{equation}
From the above relationship, an optimal offset is derived as
\begin{equation}
    {\boldsymbol{\Omega}}^*(\varphi) = \frac{1}{K}\sum_{k} \left(\hat{\mathbf z}^{(k)} -\mathbf R(\varphi)\mathbf p^{(k)}\right).
\end{equation}
Note that the optimal offset depends on the rotation angle; thus, it is expressed as a function of $\varphi$.
If we substitute the optimal offset into the original cost function~(\ref{5A4}), we have the following relationship:
\begin{align} \label{App1}
    J(\mathcal Z,& \mathcal P; \varphi, {\boldsymbol{\Omega}}^*(\varphi)) = \sum_{k} \lVert \hat{\mathbf z}^{(k)}\rVert^2 + \sum_{k} \lVert \mathbf p^{(k)}\rVert^2\nonumber\\
    -&\frac{1}{K}\lVert\sum_{k} \hat{\mathbf z}^{(k)} \rVert^2 - \frac{1}{K}\lVert\sum_{k} \mathbf p^{(k)} \rVert^2 + 2J(\varphi),
\end{align}
where $J(\varphi)$ contains all terms related to the rotation angle $\varphi$, which is given by
\begin{equation}
    J(\varphi) = \frac{\left(\sum_{k} \hat{\mathbf z}^{(k)}\right)^T\mathbf R(\varphi)\left(\sum_{k} \mathbf p^{(k)}\right)}{K} - \sum_k (\hat{\mathbf z}^{(k)})^T\mathbf R(\varphi)\mathbf p^{(k)}.
\end{equation}

Using equation~(\ref{5A2}) and symbols defined in (\ref{5A6}), the above equation is rewritten as
\begin{equation}
    J(\varphi)= \Gamma \cos\varphi + \tilde{\Gamma}\sin\varphi,
\end{equation}
and it satisfies the following inequality:
\begin{equation}
    J(\varphi) = \sqrt{\Gamma^2 + \tilde{\Gamma}^2} \cos(\varphi - \psi) \geq -\sqrt{\Gamma^2 + \tilde{\Gamma}^2},
\end{equation}
where $\psi = \arctan\frac{\Gamma}{\tilde{\Gamma}}$.
An equality condition for this inequality is given as $\varphi - \psi = \pi$.
Therefore, an optimal angle that minimizes the original cost function is expressed by
\begin{equation}
    \varphi^* = \pi + \arctan\frac{\Gamma}{\tilde{\Gamma}},
\end{equation}
and the optimal offset is determined as $\boldsymbol{\Omega}^* = \boldsymbol{\Omega}(\varphi^*)$ accordingly.
The minimum value of the cost function can be obtained from equation~(\ref{App1}) by replacing $J(\varphi)$ to $-\sqrt{\Gamma^2 + \tilde{\Gamma}^2}$.

\section{Positioning with Wi-Fi Ranging and Sensors}

To obtain the correct movement of the device from the PDR output, the reference direction should be estimated. We include unknown reference direction $\phi_{ref}$ in the EKF state as
\begin{equation}
    \boldsymbol{\zeta} = [\mathbf z^T, \phi_{ref}]^T = [x, y, \phi_{ref}]^T.
\end{equation}
The proposed EKF procedure is summarized as below.

\emph{1) Initialization}: Because the reference direction can be any direction in the x-y plane of the GCS, we simultaneously consider multiple initializations of the EKF state with different initial reference direction estimates. 
We consider $M$ candidates of the EKF state, and the $m$-th candidate is initialized as
\begin{equation}
    \hat{\boldsymbol{\zeta}}^{(0)}_m = [(\hat{\mathbf z}^{(0)})^T, \hat{\phi}_{ref, m}^{(0)}]^T,~1\leq m \leq M,
\end{equation}
where $\hat{\mathbf z}^{(0)}$ is the same as equation~(\ref{4c1}), and $\hat{\phi}_{ref, m}^{(0)}=\frac{2\pi m}{M}$ is initial reference direction estimate of the $m$-th candidate.
The covariance of the EKF state for the $m$-th candidate is initialized as
\begin{equation}
    \tilde{\mathbf P}^{(0)}_m = \mbox{diag}\left(s_x^2, s_y^2, s_\phi^2\right),
\end{equation}
where $s_\phi$ is the standard deviation of initial reference direction estimate.

\emph{2) State Prediction}: The state transition model is given by
\begin{equation}
    \boldsymbol{\zeta}^{(k)} = \tilde{\mathbf f}(\boldsymbol{\zeta}^{(k-1)}, \Delta \mathbf p^{(k)}),
\end{equation}
where $\Delta \mathbf p^{(k)} = \mathbf p^{(k)} - \mathbf p^{(k-1)}$ represents the movement of the device reported from the PDR module.
The relationship between the elements in the state is given by
\begin{equation}
    \mathbf z^{(k)} = \mathbf z^{(k-1)} + \mathbf R(\phi_{ref}^{(k-1)}) \Delta \mathbf p^{(k)},~\phi_{ref}^{(k)} = \phi_{ref}^{(k-1)}.
\end{equation}
Using the state transition model, the predicted state of the $m$-th candidate is obtained as
\begin{equation}
    \hat{\boldsymbol{\zeta}}^{(k|k-1)}_m = \tilde{\mathbf f}(\hat{\boldsymbol{\zeta}}^{(k-1)}_m, \Delta \mathbf p^{(k)}),
\end{equation}
and its covariance matrix is updated accordingly as
\begin{equation}
    \tilde{\mathbf P}^{(k|k-1)}_m = \tilde{\mathbf F}_m^{(k)} \tilde{\mathbf P}_m^{(k-1)} (\tilde{\mathbf F}_m^{(k)})^T,
\end{equation}
where $\tilde{\mathbf F}_m^{(k)} \in \mathbb{R}^{3\times 3}$ represents the Jacobian matrix defined as
\begin{equation}
    \tilde{\mathbf F}_m^{(k)} \triangleq \evalat[\Big]{\frac{\partial \tilde{\mathbf f}(\boldsymbol{\zeta}, \Delta \mathbf p^{(k)})}{\partial \boldsymbol{\zeta}}}{\boldsymbol{\zeta} = \hat{\boldsymbol{\zeta}}^{(k-1)}_m}.
\end{equation}

\emph{3) State Update}: The measurement model is expressed by
\begin{equation}
    \mathbf d^{(k)} = \tilde{\mathbf h}^{(k)}(\boldsymbol{\zeta}^{(k)}) + \boldsymbol{\omega}^{(k)} = \mathbf h^{(k)}(\mathbf z^{(k)}) + \boldsymbol{\omega}^{(k)},
\end{equation}
where $\mathbf h^{(k)}(\cdot)$ and $\boldsymbol{\omega}^{(k)}$ are defined in equation~(\ref{4c5}).
The innovation of the $m$-th candidate and its covariance matrix are given by
\begin{gather}
    \tilde{\mathbf e}^{(k)}_m = \mathbf{d}^{(k)} - \tilde{\mathbf h}^{(k)}(\hat{\boldsymbol{\zeta}}_m^{(k|k-1)}),\nonumber\\
    \tilde{\mathbf S}_m^{(k)} = \tilde{\mathbf H}_m^{(k)} \tilde{\mathbf P}_m^{(k|k-1)} (\tilde{\mathbf H}_m^{(k)})^T + \boldsymbol{\Lambda}^{(k)},
\end{gather}
respectively. Here, $\tilde{\mathbf H}^{(k)}_m \in \mathbb{R}^{N\times 3}$ represents the Jacobian matrix defined as
\begin{equation}
    \tilde{\mathbf H}^{(k)}_m \triangleq \evalat[\Big]{\frac{\partial \tilde{\mathbf h}^{(k)}(\boldsymbol{\zeta})}{\partial \boldsymbol{\zeta}}}{\boldsymbol{\zeta} = \hat{\boldsymbol{\zeta}}_m^{(k|k-1)}}.
\end{equation}
The remaining processes are similar to Section~III-D. The Kalman gain, updated state, and its covariance matrix of the $m$-th candidate are computed as
\begin{gather}
    \tilde{\mathbf{G}}_m^{(k)} = \tilde{\mathbf{P}}^{(k|k-1)}_m(\tilde{\mathbf H}_m^{(k)})^T(\tilde{\mathbf S}^{(k)}_m)^{-1},\nonumber\\
    \hat{\boldsymbol{\zeta}}^{(k)} = \hat{\boldsymbol{\zeta}}^{(k|k-1)}+\tilde{\mathbf G}_m^{(k)} \tilde{\mathbf e}_m^{(k)},\nonumber\\
    \tilde{\mathbf P}_m^{(k)} = \left(\mathbf I_3 - \tilde{\mathbf G}_m^{(k)}\tilde{\mathbf H}_m^{(k)}\right)\tilde{\mathbf P}^{(k)}_m.
\end{gather}

\emph{4) Best Candidate Selection}: Once the EKF updates the state of every candidate using the latest measurement results, the best candidate is selected based on the innovation (or accumulated innovation) of each candidate as
\begin{equation}
    m^{*} = \argmin_m \lVert \tilde{\mathbf e}_m^{(k)} \rVert^2.
\end{equation}
Then, the state of the selected candidate $\hat{\boldsymbol{\zeta}}_{m^*}^{(k)}$ is reported as the state estimate of the device at time step $k$, where the first two elements are the x- and y-coordinates, and the last element is the reference direction.

\section*{Acknowledgment}

The author would like to thank Yang-Seok Choi from Intel Labs (e-mail: yang-seok.choi@intel.com) for supporting this project and Gary Rozen (e-mail: gary.rozen@intel.com) from Intel Client Computing Group for providing the CSI tool.

\ifCLASSOPTIONcaptionsoff
  \newpage
\fi

\end{document}